\documentstyle[12pt] {article}
\newcommand{\beq}{\begin{equation}}
\newcommand{\eeq}{\end{equation}}
\newcommand{\beqa}{\begin{eqnarray}}
\newcommand{\eeqa}{\end{eqnarray}}
\newcommand{\ba}{\begin{array}}
\newcommand{\ea}{\end{array}}

\begin{document}

\begin{center}
{\bf FROM REGULAR TO CHAOTIC STATES\\
IN ATOMIC NUCLEI}\footnote{This work has been partially supported by the
Ministero dell'Universit\`a e della Ricerca Scientifica e Tecnologica
(MURST) and by the INFN--CICYT Agreement.}

\vskip 1. truecm

{\bf M.T. Lopez--Arias}$^{(+)(++)}$, {\bf V.R. Manfredi}
\footnote{Author to whom all
correspondence and reprint requests should be addressed. E--Mail:
VAXFPD::MANFREDI, MANFREDI@PADOVA.INFN.IT}
$^{(++)(*)}$\\
and {\bf L. Salasnich}$^{(+++)}$  \\

\vskip 1. truecm

(+){\it Nuclear Physics Group, Facultad de Ciencias,}\\
{\it Universidad de Salamanca, 37008 Salamanca, Spain}\\

\vskip 0.6 truecm

(++){\it Dipartimento di Fisica ``G. Galilei" dell'Universit\'a di Padova,}\\
{\it and INFN, Sezione di Padova, via Marzolo 8, 35131 Padova, Italy}\\

\vskip 0.6 truecm

(*){\it Interdisciplinary Laboratory, International School for Advanced
Studies}\\
{\it Strada Costiera 11, 34014 Trieste, Italy}\\

\vskip 0.6 truecm

(+++){\it Dipartimento di Fisica dell'Universit\'a di Firenze,}\\
{\it and INFN, Sezione di Firenze, Largo E. Fermi 2, 50125 Firenze, Italy}\\

\end{center}

\vfill \eject

\par
{\bf 1. Introduction}
\vskip 0.5 truecm
\par
An interesting aspect of nuclear dynamics is the co--existence, in atomic
nuclei, of {\it regular} and {\it chaotic} states [1]. This is a vast
subject and, for reasons of space, in this paper we limit ourselves to a
few examples only.
\par
In order to highlight the difference between the regular and chaotic states,
it is perhaps useful to remember that the low
energy states ($0\sim 4$ MeV above the ground state), termed
{\it regular}, are described by a variety of models: the
shell--model,  the collective model and their various extensions
[2--8]. By means of these models, all the properties of nuclear levels
such as excitation energies,
transition probabilities, magnetic and quadrupole momenta, etc. may be
accurately calculated.
\par
When the excitation energy increases, the level density rises,
making the calculation of nuclear properties in terms of single levels
both of no physical interest and impossible. Instead, a
statistical description, called {\it Statistical Nuclear Spectroscopy} (SNS),
is used, based on the division into {\it global} and {\it local} nuclear
properties [9--17]. A typical example of this
is the separation of the level density into a global component,
the {\it secular variation}, and a local component,
the fluctuations, which are well described by the random
matrix ensembles [18--21].
\par
In recent years, the study of quantum levels in
classically chaotic regions has shown that they have the same fluctuation
properties as those predicted by random matrix ensembles [28--30]
in a large energy range; we term these states {\it chaotic}.
The physical significance of statistical concepts in atomic nuclei
is therefore understood through the link with
chaotic motion in hamiltonian dynamics [22--25].
\par
In the first part of the present work, we review the state of the art of
nuclear dynamics and use a schematic shell model to show how a very
simple and schematic nucleon--nucleon interaction can produce an
order$\to$chaos transition. The second part is devoted to a discussion
of the wave function behaviour and decay of chaotic states using some
simple models.

\vskip 0.5 truecm

\par
{\bf 2. Regular states in atomic nuclei: shell and collective models}
\vskip 0.5 truecm
\par
As is well known, the {\it regular states} of atomic nuclei are
described, roughly speaking, by the mean field approximation of the
shell model and by the oscillations about the mean field which give rise
to collective excitations. Nuclear potential is essentially symmetric;
nucleons move in regular orbits and, besides the energy, there are other
constants of motion and other quantum numbers [6,7]. For the sake of
completeness we report below some basic notions of the {\it shell model}
and {\it collective models}.
\par
In the {\it shell model} [5] the nuclear states,
like the electronic states in atoms, are
described in terms of the motion of nucleons in a mean--field. But, as
is well known, while the nuclear field is generated by the interactions
of the nucleons,
the atomic field is mainly governed by the interaction of the electrons
with the nucleus.
\par
Experimental data suggests a shell structure for the
atomic nucleus, i.e. the greater stability of the nuclei with 2, 8, 20,
28, 50, 82 and 126 neutrons or 2, 8, 20, 28, 50, 82 protons. These
nuclei are called {\it magic nuclei} and their nucleonic numbers are
termed {\it magic numbers}. The low energy levels of these {\it magic
nuclei} are very high, so their structure is particular stable. In the
{\it shell model} the {\it magic numbers} represent the numbers of
nucleons that saturate the nuclear shells. {\it Magic nuclei} are to
nuclei what noble gases are to atoms.
\par
The hamiltonian of the {\it shell model} can be written as:
$$
{\hat H}_{0}=\sum_{i=1}^A -{\hbar^2\over 2m_i} \triangle_i
+{\hat U}_i, \eqno (2.1)
$$
where A is the number of nucleons, and ${\hat U}_i$ is the
mean--field. The choice of the mean--field is crucial; ${\hat U}_i$ may
be obtained by the usual methods using the many body theory. However,
phenomenological approximations have traditionally been used instead:
$$
{\hat U}_i={\hat U}_{C,i}(r)+{\hat U}_{LS,i}
{\vec L}_i\cdot {\vec S}_i, \eqno (2.2)
$$
where ${\hat U}_{C,i}(r)$ is the central potential, i.e. with a spherical
symmetry, and ${\hat U}_{LS,i}{\vec L}_i\cdot {\vec S}_i$ is the
spin--orbit interaction.  It is common to use a
central potential that behaves like the charge density of the
nucleus (Saxon--Woods potential):
$$
{\hat U}_{C,i}(r)={ -U_{0}\over 1+\exp{[(r-R_{0})/a]} }, \eqno (2.3)
$$
where $R_{0}$ is the nuclear radius, $a\simeq 0.53 \cdot 10^{-15}$m and
$U_{0}\simeq (50$--$60)$ MeV. With this potential the calculations are quite
complicated, so simpler potentials are usually used.
\par
To obtain a good agreement between the shell model results and the
experimental data, it is necessary to add a residual interaction ${\hat
H}_{R}$ so that the total hamiltonian ${\hat H}$ can be written:
$$
{\hat H}={\hat H}_{0}+{\hat H}_{R}, \eqno (2.4)
$$
where ${\hat H}_{R}$ is the part of nucleon--nucleon interaction
not included in ${\hat H}_{0}$. Using second quantization formalism [8],
we can write:
$$
{\hat H}_{0}=\sum_{i=1}^{A}\epsilon_{i}{\hat a}_{i}^{+}{\hat a}_{i},
\eqno (2.5a)
$$
$$
{\hat
H}_{R}=\sum_{ijkl=1}^{A}V_{ijkl}{\hat
a}_{i}^{+}{\hat a}_{j}^{+}{\hat a}_{l}{\hat a}_{k},  \eqno (2.5b)
$$
where $\epsilon_{i}$ and $V_{ijkl}$ are the single--particle energies
and the residual interactions respectively. The operators
${\hat a}_{i}^{+}$ and ${\hat a}_{i}$ are the creation and
annihilation operators of the ith single nucleon state:
$$
[{\hat a}_{i},{\hat a}_{j}^{+}]_{+}=\delta_{ij},\quad
[{\hat a}_{i},{\hat a}_{j}]_{+}= [{\hat a}_{i}^{+},{\hat
a}_{j}^{+}]_{+}=0, \eqno (2.6)
$$
where $[{\hat a},{\hat b}]_{+}={\hat a}{\hat b}+{\hat b}{\hat a}$.
${\hat H}_{0}$ and ${\hat H}_{R}$ can be calculated by the
Hartree--Fock equations [8], starting from the free nucleon interactions;
but often, as mentioned above, phenomenological approximations are used
(the values of $\epsilon_i$ are obtained by the experimental spectrum of
the nuclei with only one nucleon added to a double magic core).
\par
In order to obtain the different observables, the following procedure is
generally used. As first step we solve the unperturbed equation:
$$
{\hat H}_{0}|\phi_{\alpha}>=E_{\alpha}^{(0)}|\phi_{\alpha}>, \eqno
(2.7)
$$
where $E_{\alpha}^{(0)}=\sum_{i=1}^{A}\epsilon_{i}$.
To obtain the eigenvalues and eigenstates of ${\hat H}$,
the Schr\"odinger equation:
$$
{\hat H}|\psi_{n}>=E_{n}|\psi_{n}>, \eqno (2.8)
$$
must be solved and $|\psi_n>$ expanded by a complete unperturbed base of
eigenstates:
$$
|\psi_{n}>=\sum_{\alpha}C_{n\alpha}|\phi_{\alpha}>. \eqno (2.9)
$$
In this way we obtain the secular problem:
$$
\sum_{\beta}E_{\alpha}^{(0)} \delta_{\alpha \beta}+<\phi_{\alpha}
|{\hat H}_{R}|\phi_{\beta}>C_{n\beta}=E_{n}C_{n\alpha}. \eqno (2.10)
$$
It is clear that the sums in (2.9) and (2.10) are
infinite and so the secular problem is not solvable.
To overcome this difficulty, it is standard procedure to cut the basis
states by introducing a
finite number of configurations which are sufficient to describe
the first excitation states. Many nucleons are frozen in the
deeper shells of the mean field potential and form an inert
core; only a few nucleons partially populate the single
particle shells outside the core. These are called valence--nucleons.
So there are N valence--nucleons, m active shells and a finite
number of energy levels.
\par
Now, if we know the states $|\psi_{n}>$, it is possible to calculate
all the observables A, which characterize a nuclear state. To obtain
the value of A in the state $|\psi_{n}>$, we must calculate the
diagonal element $<\phi_{n}|{\hat A}|\phi_{n}>$ of the operator
${\hat A}$ associated to A:
$$
A_{\psi_{n}}=<\psi_{n}|{\hat A}|\psi_{n}>=\sum_{\alpha,\beta}
C_{n\alpha}C_{n\beta}<\phi_{\alpha}|{\hat A}|\phi_{\beta}>.
\eqno (2.11)
$$
The non--diagonal elements $<\psi_{f}|{\hat A}|\psi_{i}>$ are
associated with transition probabilities between
the initial states $|\psi_{i}>$ and the final states $|\psi_{f}>$.
\par
If the number of valence--nucleons is high, this procedure becomes
very complicated; on the other hand, the low energy spectrum of nuclei
with many nucleons outside the closed shells shows a simple
behaviour which changes systematically from one nucleus to
another. These regularities are explained by describing the
nuclear correlations with {\it collective motions}, corresponding to
variations in the shape of the nucleus. In this way, we obtain a
generalization of the shell model: the mean field is not an isotropic
static potential, but becomes a variable field that can assume various
shapes, not only with spherical symmetry, but also, for example,
ellipsoidal. For these nuclei the {\it collective motions} can be divided
into {\it rotations} and {\it vibrations} [6,7]. The first correspond to
the rotation of the nuclear orientation with shape conservation and
have low excitation energies; the second correspond to oscillations
of the nucleus around its equilibrium shape.

\par
The nuclear surface in polar coordinates can be written:
$$
R(\theta ,\phi )=R_{0} \{ 1+\sum_{\lambda
=0}^{\infty}\sum_{\mu=-\lambda}^{\lambda}\alpha_{\lambda \mu}
Y_{\lambda \mu}(\theta ,\phi ) \} = R_{0}+\Delta R, \eqno (2.12)
$$
where $R_{0}$ is the mean nuclear radius, $Y_{\lambda \mu}
(\theta ,\phi)$ are the spherical harmonics and
$\alpha_{\lambda \mu}$ are the coefficients which describe the
deformation, generally time--dependent.
Small oscillations around the equilibrium shape
may be described by harmonic oscillators, so the hamiltonian of
the system can be written:
$$
{\hat H}_{\lambda}={B_{\lambda}\over2}\sum_{\mu=-\lambda}^{\lambda}
|{\dot \alpha_{\lambda \mu}}|^{2}+{C_{\lambda}\over 2}
\sum_{\mu=-\lambda}^{\lambda}|\alpha_{\lambda \mu}|^{2}, \eqno (2.13)
$$
where the coefficients $B_{\lambda}$ and $C_{\lambda}$ are related
to the frequency vibration $\omega_{\lambda}$ by:
$$
\omega_{\lambda}=\sqrt{{B_{\lambda}\over C_{\lambda}}}. \eqno (2.14)
$$
If we consider the low energy spectrum ($0\sim 4$ MeV), only $\lambda
=2,3$ are important, because $\lambda =0$ represents a compression (or
dilatation) without shape change and $\lambda =1$ represents a
translation of the entire nucleus or a dipole oscillation. Both modes
are outside the energy range considered. So each quantum excitation,
called {\it phonon}, has an energy $E_{\lambda}=\hbar
\omega_{\lambda}$ and spin $I=\lambda \hbar$.
\par
For deformed nuclei, the rotation excitation
energies may be written:
$$
E_{I,k}={\hbar^2\over 2 J}[I(I+1)-k^2]+{\hbar^2\over 2J_{3}}k^2,
\eqno (2.15)
$$
where $I$ is the total angular momentum of the nucleus and $k$ the
projection of $I$ on the symmetry axis. $J (=J_{1}=J_{2})$ and
$J_{3}$ are the momenta of inertia of the nucleus referred to the
principal axes. Equation (2.15) becomes simpler in the case of
the k=0 rotational band:
$$
E_{I}=  {\hbar^2\over 2 J}I(I+1), \eqno (2.16)
$$
where the allowed values for the angular momentum $I$ are 0, 2, 4,
6, ... . For low angular momenta ($I < 8\hbar$),
there is a good agreement between the energy values of the {\it
rotational model} and the experimental data.

\vskip 0.5 truecm
{\bf 3. Chaotic states in atomic nuclei}
\vskip 0.5 truecm
\par
In the low energy excitation spectrum of an atomic nucleus, the average
level density $\bar{\rho}(E)$ is small and, as discussed in the previous
section, one might expect to be able to describe most of the states
in detail using nuclear models. However, the average level density
increases very rapidly with the excitation energy $E$ (Bethe's law):
$$
\bar{\rho}(E) ={C\over (E-\Delta )^{5\over 4}} \exp{(A\sqrt{E-\Delta})},
\eqno (3.1)
$$
where $A$,$C$,$\Delta$ are constants for a given nucleus [2].
Therefore, once the region of the neutron emission threshold is reached
($\sim 8$ MeV), the number of levels is so high that a description
of the individual levels has to be abandoned.
The aim of nuclear models at
this and higher excitation energies is rather to describe special states, like
giant resonances and other collective states, which have a peculiar structure.
But the detailed description of the sea of background states around the
collective ones is fruitless. Thus, for example, observations of
levels of heavy nuclei in the neutron--capture region give precise information
concerning a sequence of levels from number N to number (N + n), where N is
an integer of the order of $10^6$ (see fig. 1) [16].
For these densities of states a level assignment based on shell or
collective models becomes a very difficult task.
It is therefore reasonable to inquire whether the
highly excited states may be understood from the opposite point
of view, assuming as working hypothesis that the shell structure is
completely washed out and that no quantum numbers other than spin and
parity remain good. The outcome of such an approach is termed a
{\it statistical theory of energy levels} [10--12].
\par
The statistical theory will not predict the detailed sequence of levels
in any one nucleus, but it will describe the general appearance and the
degree of irregularity of the level structure that is expected to occur
in the atomic nuclei, which are too complicated to be understood in
detail.
\par
We describe a complex nucleus as a "black box", in which a large number
of particles are interacting according to unknown laws [13]. The problem
is then to define, in a mathematically precise way, an ensemble
in which all possible laws of interaction are equally probable. This
program, initiated by Wigner and developed by many authors, has,
to a large extent, been successful [13] .

\vspace{0.5 cm}
{\bf 3.1 The random matrix theory}
\vspace{0.5 cm}

An appropriate means to define an ensemble of random matrices is
provided by the {\it random matrix theory} [14,15]. The hamiltonian
matrix $H$ is an NxN stochastic matrix (its matrix elements
$h_{ij}$ are random variables) and the probability density is
specified by:
$$
P(H)dH=P(h_{11},h_{12},...,h_{NN})
dh_{11}dh_{12}...dh_{NN}, \eqno (3.2)
$$
where the {\it probability information} is:
$$
I\{P(H)\}=\int dH P(H)\ln{P_{N}(H)}. \eqno (3.3)
$$
The aim is to find the function $P(H)$ that minimizes $I$. This is
equivalent to assuming the least possible knowledge about $P(H)$.
We impose that the $h_{ij}$ are real and, to limit the
eigenvalues of $H$ to a finite range, a condition on its norm
$[Tr(H^2)]^{1/2}$ is also imposed. Thence $P(H)$ should minimize
$I$ subject to the constraints:
$$
\int dH P(H)=1, \int dH P(H) Tr(H^2) =C, \eqno (3.4)
$$
which leads to:
$$
P(H)=\exp{\{ \lambda_{1} +\lambda_{2}Tr( H^2) \} }. \eqno (3.5)
$$
By inserting (3.5) in (3.4), the Lagrange multipliers
$\lambda_{1},\lambda_{2}$ may be determined. This probability
distribution defines the so--called {\it Gaussian Orthogonal Ensemble}
(GOE), since $P(H)$ is invariant for orthogonal transformation and
the elements $h_{ij}$ are independent random variables.
We can observe that, without the time invariance, the $h_{ij}$ are
complex numbers and we have the {\it Gaussian Unitary Ensemble} [14].

With the help of (3.5) one can easily obtain
$P(E)=P(E_{1},E_{2},...,E_{N})$ and so the {\it statistic} of the
energy spectrum. Mehta defines "{\it statistic}" thus:
``{\it A statistic is a quantity which can be
calculated from an observed sequence of levels alone, without other
information and whose average value and variance are known from the
theoretical model. A suitable statistic is one which is sensitive
for  the property to be compared or distinguished and is insensitive
for other details}" [15]. Various statistics may be used to show the
local correlations of the energy levels; in this work we shall use
$P(s)$ and $\Delta_{3}(L)$ mainly.

The first statistic measures the probability that two neighboring
eigenvalues are a distance ``s" apart, in the average level distance
unit.  For this statistic the GOE average is closely approximated by
the Wigner distribution:
$$
P(s)={\pi \over 2}s\exp{(-{\pi \over 4}s^2)}, \eqno (3.6)
$$
which gives level repulsion.

The second statistic is defined for a fixed interval $(-L/2,L/2)$, as
the least--square deviation of the staircase function $N(E)$ from
the best straight line fitting it:
$$
\Delta_{3}(L)={1\over L}\min_{A,B}\int_{-L/2}^{L/2}[N(E)-AE-B]^2 dE,
\eqno (3.7)
$$
where $N(E)$ is the number of levels between E and zero for positive
energy, between $-E$ and zero for negative energy. The $\Delta_{3}(L)$
statistic provides a measure of the degree of rigidity of the
spectrum: for a given interval L, the smaller $\Delta_{3}(L)$ is,
the stronger is the rigidity, signifying the long--range
correlations between levels. For this statistic in the GOE ensemble:
$$
\Delta_{3}(L)=\cases{{L\over 15}, \quad L \ll 1 \cr
              \noalign{\vskip 16 truept}
            {1\over \pi^2}\log{L},\quad L \gg 1 \cr }.\eqno (3.8)
$$
For the GUE there are similar results [14].

If the mean level density for the GOE is calculated, we obtain:
$$
\rho(E)=\cases{{1\over \pi \sigma^2}\sqrt{2N\sigma^2-E},
               \quad |E|<2N\sigma^2 \cr
              \noalign{\vskip 16 truept}
               0, \quad |E|<2N\sigma^2  \cr }. \eqno (3.9)
$$
Although this is an unrealistic result, it can be explained by
remembering that global and local behaviours are on different scales
and GOE is a good model for local properties.

\vspace{0.5 cm}
{\bf 3.2 Comparison of GOE predictions with experimental data}
\vspace{0.5 cm}

Neutron resonance spectroscopy on a heavy even--even nucleus
typically leads to the identification of about 150 to 170 s--wave
resonances with $J^{\pi}={1\over 2}^{+}$ located 8--10
MeV above the ground state of the compound system, with average
spacings around 10 eV and average total widths around 1 eV. Proton
resonance spectroscopy yields somewhat shorter sequences of levels
with fixed spin and parity, with typically 60 to 80 members. For
the statistical analysis, it is essential that the sequences be pure
(no admixture of levels with different spin or parity) and
complete (no missing levels) [17]. Only such sequences were considered
by Haq, Pandey and Bohigas [18]. Scaling each sequence to the same
average level spacing and lumping together all sequences one
leads to the ''Nuclear
Data Ensemble" (NDE), which contains 1726 level spacings. As shown
in fig. 2 the agreement between the experimental data and the
GOE predictions is surprisingly good (in the GOE model there are no
free parameters) [19].

It is also important to observe that this agreement in not limited to nuclear
levels, but is valid for all quantum systems (nuclear, atomic and
molecular) in a high energy range [1,29].

The behaviour of spectral statistics near the ground state is also of
considerable interest. Garrett, German, Courtney and Espino [20] (fig. 3)
and also Shriner, Mitchell and Von Egidy [21] (fig. 4) have shown that
for these energies nuclei do not follow the GOE results but behave
like a Poisson ensemble or with an intermediate behaviour
between Poisson and GOE. In the Poisson ensemble:
$$
P(s)=\exp{(-s)}, \quad \Delta_{3}(L)={L\over 15}, \eqno (3.10)
$$
which gives a high probability for the occurrence of near
degeneracies, no level repulsion and no correlation between
levels.

To sum up, we have seen that the spectra of atomic nuclei show a
Poisson$\to$GOE transition on increasing the excitation energy
(see fig. 5) over the yrast line.
In the next section we show how this behaviour is not of a
purely quantal nature but has a classical counterpart in
the regular and chaotic hamiltonian systems.

\vspace{0.5 cm}
{\bf 4. Order and chaos in classical and quantum mechanics}
\vspace{0.5 cm}

In this section we introduce the dynamical systems and their
stability, which may be studied by means of
the Lyapunov exponents and metric entropy. With the help of these
quantities, we clarify the concept of ergodic system
giving a hierarchy of chaos. Then we extend the study to hamiltonian
systems, which are essential in order to study the transition
order$\to$chaos in nuclear physics.

A {\it dynamical system} [22--25] is defined by N differential
equations of the first order:
$$
{d\over dt}{\vec z}(t)={\vec f} ({\vec z}(t),t), \eqno (4.1)
$$
where the variables ${\vec z}=(z_{1},...,z_{N})$ are in the phase
space $\Omega$ (the euclidean space $R^{N}$, unless otherwise
specified). These equations describe the time evolution of the
variables and the system they represent.

A solution of the dynamical system is a vector function
${\vec z}({\vec z}_{0},t)$, that satisfies (4.1) and the initial
condition:
$$
{\vec z}({\vec z}_{0},0)={\vec z}_{0}, \eqno (4.2)
$$
often written simply ${\vec z}(t)$ without the initial condition
dependence.

The time evolution of ${\vec z} \in \Omega$ is obtained with the one
parameter group of diffeomorfism $g^{t}$: $\Omega \to \Omega$, so that:
$$
{d\over dt}(g^{t}{\vec z})|_{t=0}={\vec f} ({\vec z},0). \eqno (4.3)
$$
The group $g^{t}$ is called {\it phase flux} and the solution is
called {\it orbit}. The system is called {\it hamiltonian},
if the dimension of $\Omega$ is even and there exists a function
$H({\vec z},t)$ given by:
$$
{\vec f} ({\vec z}(t),t)=J\nabla H({\vec z},t), \eqno (4.4)
$$
where:
$$
J= \left( \matrix {0 & I \cr -I & 0 \cr } \right) \eqno (4.5)
$$
is the symplectic matrix. $H({\vec z},t)$ is the {\it hamiltonian}.

On the phase space $\Omega$ one usually defines a probability measure
$\mu : \Omega \to \Omega$, so that $\mu (\Omega )=1$.
If we choose a subspace A of
$\Omega$, the system is measure preserving if:
$$
\mu (g^{t}A)=\mu (A). \eqno (4.6)
$$
It is well known that hamiltonian systems preserve their measure:
the Liouville measure. Dynamical systems which do not preserve their
measure are called {\it dissipative}, and usually have
a measure contraction in time evolution.

The dynamic of a system is called {\it regular} if the orbits are
stable to infinitesimal variations of initial conditions. It
is called {\it chaotic} if the orbits are
unstable to infinitesimal variations of initial
conditions [23--25]. Useful quantities to calculate this behaviour are the
Lyapunov exponents, which give the stability of a single orbit, and
the metric entropy, which represents a mean exponent for the entire
system.
A vector of the tangent space $T\Omega_{\vec z}$ to the phase space
$\Omega$ in the position ${\vec z}$ is given by:
$$
{\vec \omega} ({\vec z})=\lim_{s \to 0} {{\vec q}(s)-{\vec q}(0) \over s},
\eqno (4.7)
$$
where ${\vec q}(0)={\vec z}$ and ${\vec q}(s) \in \Omega$. The tangent
space vectors are the velocity vectors of the curves on M; there are
obviously N independent vectors.

Now we can define the Lyapunov exponent:
$$
\lambda ({\vec z})=\lim_{t\to \infty}{1\over t}\ln{|\omega (t)|},
\eqno (4.8)
$$
where ${\vec \omega} (t)$ is a tangent vector to ${\vec z}(t)$ with the
condition that $|{\vec \omega} (0)|=1$.

It can be demonstrated that the limit given by (4.8) exists for a
compact phase space, and that it is metric independent. Fixing an
orbit in the N dimensional phase space, there are N distinct
exponents $\lambda_{1},...,\lambda_{N}$, called first order
Lyapunov exponents. If the orbit has positive Lyapunov exponents, it
is chaotic.

To characterize globally the chaoticity of a system, we introduce
the metric entropy. If $\alpha (0)$ is a partition of
non--overlapping sets that completely cover the phase space $\Omega$ at
the initial instant:
$$ \alpha (0)=\{A_{i}(0): \Sigma (E)=\cup_{i}A_{i}(0),A_{i}(0)\cap
A_{j}(0)=\phi \},  \eqno (4.9)
$$
the partition $\alpha (0)$ can evolve in a discretized time flux:
$$
\alpha (0),\alpha (1),\alpha (2),...,\alpha (n). \eqno (4.10)
$$
Then we define $\beta (n)$ as the intersection set of all the sets
at every instant:
$$
\beta (n)=\{ B_{l}(n): B_{l}(n)=A_{i}(0)\cap A_{j}(1)\cap ...\cap
A_{k}(n)\}. \eqno (4.11)
$$
In this way, the number of sets defined by $\beta (n)$ do not
decrease when $n$ is increased. By introducing a probability measure $\mu$
preserved by dynamics, one can define the metric information of the
partition $B(n)$:
$$
I_{n}=-\sum_{i} \mu (B_{i}(n))\ln{\mu (B_{i}(n))}. \eqno (4.12)
$$
The information $I_{n}$ has the following properties:

(i) $I_{n}=0$ if, and only if, there exists a set $B_{i}(n)$ of the
partition $\beta (n)$ so that $\mu (B_{i}(n))=1$;

(ii) $I_{n}$ assumes the maximum value when $\beta (n)=\{
B_{1}(n),...,B_{N}(n)\}$ and in this way $\mu (B_{i}(n))={1\over
N}\quad   \forall i$ i.e. $I_{n}=\ln{N}$.

The metric entropy of the partition $\alpha (0)$ is defined as:
$$
h(A_{i}(0),\mu )=\lim_{n\to \infty}{I_{n}\over n} \eqno (4.13)
$$
i.e. a time average of the metric information. If all the elements
of $\alpha (t)$ have a measure that decreases, on
average, exponentially with t, then the metric entropy will be
positive. A positive entropy indicates that there does not exist a
finite number of measures to guess the next one.

We indicate the maximum value of $h$ on all partitions $A_{i}$ with
$h_{KS}$:
$$
h_{KS}=max \{h(A_{i}(0),\mu ),\forall A_{i}(0) \}, \eqno (4.14)
$$
called {\it Kolmogorov--Sinai entropy}.
According to a very important theorem [26]:
$$
h_{KS}(\mu )=\int_{A} d\mu ({\vec z}) \sum_{\lambda_{i}>0}
\lambda_{i}({\vec z}) \eqno (4.15)
$$
with $A$ subspace of $\Omega$ and $\lambda_{i}$ Lyapunov exponents.
Therefore the {\it Kolmogorov--Sinai entropy} is a very useful tool
for showing chaotic behaviour in the region $A$.

A system is called {\it ergodic} if the time average is equal to
phase space average:
$$
\lim_{t\to\infty}{1\over t}\int_{0}^{t} dt f(g^{t}{\vec z}(t))=
\int_{\Omega} d\mu ({\vec z}) f({\vec z}). \eqno (4.16)
$$
Incidentally, as is well known, Boltzmann started from the ``ergodic
hypothesis" to obtain statistical mechanics of equilibrium.
But ergodicity is not sufficient to reach an equilibrium state:
one must consider {\it mixing systems}.

In a {\it mixing system}, every finite element of the phase space
occupies for $t\to \infty$ the entire phase space $\Omega$;
more precisely:
$\forall A,B \subset \Omega$ with $\mu(A)$ and $\mu (B) \not= \phi$,
$$
\lim_{t \to \infty} {\mu (B \cap g^{t}A)\over \mu (B)}=\mu (A).
\eqno (4.17)
$$
To have quantitative information of orbit separations, we must
introduce {\it K--systems} (Kolmogorov), which are mixing systems
with a positive metric entropy:
$$
h_{KS}>0. \eqno (4.19)
$$
Such systems are typical chaotic systems.

Among the K--systems, the most unpredictable ones are the
{\it B--systems} (Bernoulli), which have the Kolmogorov--Sinai
entropy equal to the entropy of every partition:
$$
h_{KS}=h(A_{i}(0),\mu ), \quad \forall A_{i}(0). \eqno (4.20)
$$

\vspace{0.5 cm}
{\bf 4.1 Classical chaos}
\vspace{0.5 cm}

In classical mechanics, the state of a system of coordinates
$q_{i}$ and moments $p_{i}$, $i=1,...,n$ in the N=2n dimensional
phase space $\Omega$, is specified by the hamiltonian $H=H({\vec p},
{\vec q})$, with ${\vec p}=(p_{1},...,p_{n})$, ${\vec q}=
(q_{1},...,q_{n})$ [22].

As is well known, the time evolution is obtained by the Hamilton
equations:
$$
{\dot q_{i}}={\partial H \over \partial p_{i}}, \;\;\;
{\dot p_{i}}=-{\partial H \over \partial q_{i}}.
\eqno (4.21)
$$
These equations, with the position ${\vec z}=(q_{1},...,q_{n},
p_{1},...,p_{n})$, can be written in the more compact form (4.4).

The hamiltonian system is integrable if there are n functions
defined on $\Omega$:
$$
F_{i}=F_{i}({\vec z}) \quad i=1,...,N \eqno (4.22)
$$
in involution:
$$
[ F_{i},F_{j} ]_{PB}=\sum_{k=1}^{n}
{\partial F_i\over \partial q_k}{\partial F_j\over \partial p_k}-
{\partial F_j\over \partial q_k}{\partial F_i\over \partial p_k}
=0, \quad \forall i,j \eqno (4.23)
$$
and linearly independent. $[\; ,\; ]_{PB}$ are the Poisson Brackets.

For the conservative systems we have $F_{1}=H({\vec z})$ and also:
$$
{dF_{i}\over dt}=[H,F_{i}]_{PB}=0. \eqno (4.24)
$$
Because there are n constants of motion, every orbit can explore only the n
dimensional manifold $\Omega_{f}$:
$$
\Omega_{f}=\{{\vec z}: F_{i}({\vec z})=f_{i}, i=1,...,N\} \eqno (4.25)
$$
If $\Omega_{f}$ is compact and connected, it is equivalent to a n
dimensional torus:
$$
T^{n}=\{ (\theta_{1},...,\theta_{n})\quad mod\quad 2\pi \}
\eqno (4.26)
$$
There are n irreducible and independent circuits $\gamma_{i}$ on
$\Omega_{f}$ and there exists a canonical transformation:
$$
({\vec p},{\vec q}) \to ({\vec I},{\vec \theta}) \eqno (4.26)
$$
generated by the function $S({\vec q},{\vec I})$, so that:
$$
I_{i}=\oint_{\gamma_{i}} d{\vec q}\cdot{\vec p}, \;\;\;
\theta_{i}={\partial S\over \partial I_{i}}. \eqno (4.27)
$$

The $I_{i}$ are called action variables and the $\theta_{i}$ are
called angle variables.
The moments ${\vec p}$ and coordinates ${\vec q}$ are periodic
functions of ${\vec \theta}$ with period $2\pi$:
$$
{\vec q}=\sum_{{\vec m}}{\vec q}_{{\vec m}}({\vec I}) e^{i{\vec m}
\cdot{\vec \theta}},
\;\;\;
{\vec p}=\sum_{{\vec m}}{\vec p}_{{\vec m}}({\vec I}) e^{i{\vec m}
\cdot{\vec \theta}}
\eqno (4.28)
$$
where ${\vec m}=(m_{1},...,m_{n})$ is an integer vector.

The hamiltonian depends only on action variables: $H=H({\vec I})$,
and so the new equation of motion are:
$$
{\dot\theta}_{i}={\partial H({\vec I}) \over \partial I_{i} }=
\omega_i ({\vec I}), \;\;\;
{\dot I}_{i}=-{\partial H({\vec I}) \over \partial \theta_{i} }=0,
\eqno (4.29)
$$
so:
$$
\theta_{i}=\omega_i({\vec I})t+\alpha_{i}, \eqno (4.30)
$$
where $\alpha_{i}$ are constants and
$\omega_{i}$ are the frequencies on the torus.

The orbits of an integrable system are quasi--periodic, with n
quasi--periods:
$$
T_{i}={2\pi \over \omega_{i}}, \eqno (4.31)
$$
and the orbit is closed if there exists a period $\tau$ so that:
$$
{\vec \theta} (\tau )={\vec \theta} (0)+2 \pi {\vec D}, \eqno (4.32)
$$
with ${\vec D}$ integer vector.
To have the closure, we must have:
$$
{\vec \omega}\cdot {\vec m}_{i} = 0, \eqno (4.33)
$$
and the period is:
$$
\tau = {2\pi \over \omega_{c}}={2\pi D_{i}\over \omega_{i}}={D_{i}\over
T_{i}} . \eqno (4.34)
$$
If we do not have n independent relations among frequencies, the orbits
do non close: we have the so called "torus ergodicity", but the motion
is still regular (see fig. 6).

Adding a small perturbation $V({\vec I},{\vec \theta})$
to an integrable hamiltonian $H_0({\vec I})$, the total hamiltonian can
be written:
$$
H({\vec I},{\vec \theta})=H_{0}({\vec I})+\chi V({\vec I},{\vec \theta}),
\eqno (4.35)
$$
and, generically, the integrability is destroyed. As a consequence,
parts of phase space become
filled with chaotic orbits, while in other parts the toroidal
surfaces of the integrable system are deformed but not destroyed; thus
we have a {\it quasi--integrable system}.

With growing $\chi$, chaotic motion develops near the
regions of phase space where all the $\omega_{i}$ are
commensurate; or, more precisely, where the equation (4.33) is
obeyed by integer vectors.

Conversely, tori of the integrable system on which the $\omega_{i}$
are incommensurate are deformed, but not destroyed immediately (KAM
theorem). As $\chi$ increases, the phase space generically
develops a highly complex structure, with islands of regular
motion (filled with quasi--periodic orbits)
interspersed in regions of chaotic motion, but containing in turn
more regions of chaos. As $\chi$ grows further, the
fraction of phase space filled with chaotic orbits grows until it
reaches unity as the last KAM surface is destroyed. Then the
motion is completely chaotic everywhere, except possibly for
isolated periodic orbits [22,23].

It is very useful to plot a $2n-1$ surface of section ${\cal P}
\subset \Omega$, called {\it Poincar\'e section} (see fig. 7).
As shown in figure 8, for an integrable system with two degrees of
freedom, the $x=0$ Poincar\'e section of a rational (resonant)
torus is a finite number of points along a closed curve,
while the section of an irrational
(non resonant) torus is a continuous closed curve.

Adding a perturbation, the section presents closed curves (KAM
tori), whose points are stable (elliptic), and also curves formed
by substructures, residua of resonant tori, whose points are
unstable (hyperbolic). As the perturbation parameter increases, the
closed curves are distorted and reduced in number.

\vspace{0.5 cm}

{\bf 4.2 Quantum chaos}
\vspace{0.5 cm}

In non relativistic quantum mechanics, the time evolution of a state
$|\psi >$ of a system with hamiltonian ${\hat H}$ is given by the
Schr\"odinger equation [27]:
$$
i\hbar {\partial\over \partial t}|\psi >={\hat H}|\psi >. \eqno (4.36)
$$
The unitary time--evolution operator ${\hat U}(t)$ is:
$$
{\hat U}(t)=\exp{(-{i{\hat H}\over \hbar})}, \eqno (4.37)
$$
so that $|\psi (t)>={\hat U}(t) |\psi>$. If we choose a new state
$|\psi^{'}>$ near to $|\psi>$ we have:
$$
<\psi^{'} (t) |\psi (t)>=<\psi^{'}|{\hat U}^{+}(t){\hat U}(t) |\psi (t)>=
<\psi^{'}|\psi>. \eqno (4.38)
$$
The distance between two quantum states is always time-independent, so there is
no chaos in quantum mechanics in the sense of {\it time exponential divergence
of near states}. However an interesting question is: {\it whether there
are quantum--mechanical manifestations of classical chaotic motion}.
We shall use the term {\it quantum chaotic system} in the precise, and
restricted, sense of a quantum system whose classical analogue is chaotic
[24,25].

The studies of the eigenvalues of billiards with different shapes [28,29]
have shown that if the system is classically integrable the spectral
statistics are well
modelled by the Poisson ensemble (3.10) and if the system is classically
chaotic, by the GOE (3.6, 3.8) or GUE ensembles,
depending on the time--reversal symmetry (fig. 9).

When the classical dynamics of a physical system is regular,
the short--range properties of the corresponding quantal
spectrum tend to resemble those of a spectrum of randomly distributed
numbers (the Poisson spectrum). This is because regular classical motion is
associated with integrability or separability of the classical equations of
motion.
In quantum mechanics the separability corresponds to a number of
independent conserved quantities (such as angular momentum), and each
energy level can be characterised by the associated quantum numbers.
Superimposing the terms arising from the
various quantum numbers, a spectrum is generated like that of
random numbers, at least over short intervals.
When the classical dynamics of a physical system is chaotic,
the system cannot be integrable
and there must be fewer constants of motion than degrees of freedom.
Quantum mechanically this means that once all good quantum numbers due to
obvious symmetries etc. are accounted for, the energy levels cannot simply
be labelled by quantum numbers associated with certain constants of motion.
The short--range properties of the energy spectrum then tend to resemble
those of eigenvalue spectra of matrices with randomly chosen elements.

Berry and Tabor [30] predict level clustering for any integrable system
(with the exception of the harmonic oscillator that has equal spaced
levels [31]) in the asymptotic high energy regime, using the semiclassical
($\hbar \to 0$) Einstein--Brillowin--Keller (EKB) quantization of action
variables [24]:
$$
E_{n_1,...,n_N}=H(I_1=\hbar (n_1+a_1/4),...,I_N=\hbar (n_N+a_N/4)),
\eqno (4.39)
$$
where $a_l$ is the Maslov index for the $I_l$ action variable.
In one dimension: $a =0$ for the rotator and $a = {1\over 2}$
for the oscillator.

In quasi-integrable and chaotic systems the EKB quantization is not
applicable but we can calculate the level density $\rho (E)$ using a
semiclassical formula obtained by Gutzwiller [32,33]:
$$
\rho (E)={\bar \rho} (E)+\rho_{osc}(E), \eqno (4.40)
$$
where:
$$
{\bar \rho}(E)={1\over (2\pi\hbar)^{N}}\int d{\vec p}d{\vec q} \delta (E-
H({\vec p},{\vec q})), \eqno (4.41)
$$
$$
\rho_{osc}(E)=\sum_{j}{T_{j}\over 2 f(\lambda_{j})}
\cos{({S_{j}(E)\over \hbar}-{a_{j}\pi\over 2})}. \eqno (4.42)
$$
The summation is over all the periodic orbits of the classical phase
space, $T_{j}$ is the j-periodic orbit period,
$f$ is a function of the j--periodic orbits Lyapunov exponent and
$S_{j}(E)=\oint_{j} d{\vec q}\cdot {\vec p}$.

With the help of (4.42) Berry [34] has calculated the semiclassical
behaviour of
the spectral rigidity $\Delta_{3}(L)$ for a chaotic time--reversal system:
$$
\Delta_{3}(L)=\cases{{L\over 15}, \quad L \ll 1 \cr
              \noalign{\vskip 16 truept}
            {1\over \pi^2}\log{L},\quad 1 \ll L \ll L_{max} \cr
              \noalign{\vskip 16 truept}
      {1\over 2\pi^2}\log{eL_{max}}+oscillations, \quad L \gg L_{max} \cr},
\eqno (4.43)
$$
where $L_{max}={\hbar {\bar \rho}\over T_{min}}$ with $T_{min}$ the period
of the shortest periodic orbits of the classical phase space.
The long--range periodic orbits contribute to the universal behaviour,
but there is also a {\it non universal behaviour} not predicted by GOE.

In the next section, we describe a schematic shell model to show how an
order to chaos transition may occur by studying the classical
and quantum properties of the model.

\vspace{0.5 cm}

{\bf 5. Order to chaos transition in a schematic shell model}
\vspace{0.5 cm}

To study the transition from regular to chaotic
states in atomic nuclei,
an extension of the two level Lipkin--Meshkov--Glick (LMG) model
was proposed by Meredith, Koonin and Zirnbauer (MKZ) [35--37].
The two level LMG model has only
one degree of freedom, i.e. the particle number in the upper level, and so
its classical limit does not have chaotic behaviour. The generalization
proposed by MKZ has the SU(3) symmetry and two degrees of freedom: its
classical limit shows an order to chaos transition.

The SU(3) model consists of M identical particles,
labelled by the index n, each of which can be in three
single--particle states having energy $\epsilon_{i}$. The
hamiltonian is:
$$
{\hat H}=\sum_{i=1}^3 \epsilon_{i}{\hat G}_{ii} + \sum_{ij=1}^3 V_{ij}
{\hat G}_{ij}^2, \eqno (5.1)
$$
with $\epsilon_3=\epsilon_1=\epsilon$ and $\epsilon_2 =0$,
where $V_{ij}$ is the strength of the two--body interaction between
states i and j:
$$
V_{ij}=V_{ji} \quad and \quad V_{ii}=0, \eqno (5.2)
$$
with:
$$
{\hat G}_{ij}=\sum_{n=1}^M {\hat a}_{ni}^{+} {\hat a}_{nj}. \eqno (5.3)
$$
The operator ${\hat a}_{ni}^{+}$ creates a particle n in state i, and
${\hat a}_{ni}$ annihilates a particle n from state i; each term of
${\hat G}_{ij}$ takes a particle, n, out of state j and into state i, and
each ${\hat G}_{ii}$ counts the particles in state i.

Applying the usual fermion anti--commutation rules to the
${\hat a}_{ni}$ operators, we get:
$$
[{\hat G}_{ij},{\hat G}_{hk}]_{-}={\hat G}_{ik}\delta_{jh}-{\hat G}_{jh}
\delta_{ik}, \eqno (5.4)
$$
and the 9 operators ${\hat G}_{ij}$ are generators for U(3).
Taking into account the number conservation, $\sum_{i=1}^3 {\hat G}_{ii} =M$,
this becomes SU(3).

\vspace{0.5 cm}
{\bf 5.1 Classical limit}
\vspace{0.5 cm}

The quantum 4--dimensional phase space of SU(3) is $U(3)/U(2)\otimes
U(1)$, where $U(2)\otimes U(1)$ is the maximal stability subgroup which leaves
the ground state $|00>$ invariant up to a phase factor [38].

The coherent states of $U(3)/U(2)\otimes U(1)$ are:
$$
|SU(3),{\hat \Phi} > ={\hat \Phi} |00>, \eqno (5.5)
$$
where
$$
{\hat \Phi} = \exp{\{ \sum_{i=2}^3 (\nu_{i} {\hat G}_{i1} - \nu_{i}^{*}
{\hat G}_{1i}) \} }, \eqno (5.6)
$$
with $\nu_{i}$ complex numbers.

The Bergmann kernel is:
$$
k({\vec z},{\vec z}^{*})=<SU(3),{\hat \Phi}|SU(3),{\hat \Phi}>
=(1+Z^{+}Z)^{M},
\eqno (5.7)
$$
where $Z$ is the 3x1 matrix:
$$
Z=(z_{i})=(\nu_{i}) {tan(\nu ) \over \nu}, \eqno (5.8)
$$
with $\nu=\sqrt{ \nu_{2}\nu_{2}^{*}+\nu_{3}\nu_{3}^{*}}$.
The metric of $U(3)/U(2)\otimes U(1)$ is then given by:
$$
g_{ij}={\partial^2 \ln{k({\vec z},{\vec z}^{*})} \over \partial z_{i}
\partial z_{i}^{*} }, \eqno (5.9)
$$
and the nondegenerate closed two--form $\omega$ of $U(3)/U(2)\otimes
U(1)$ is:
$$
\omega =\sum_{i,j=2}^3 g_{ij} dz_{i}\wedge dz_{j}^{*}, \eqno (5.10)
$$
with Poisson Brackets:
$$
[ f,g ]_{PB}=-i\sum_{i,j=2}^3 g^{ij}[{\partial f \over \partial z_{i}}
{\partial g \over \partial z_{j}^{*}}-{\partial f \over \partial
z_{j}^{*}}{\partial g \over \partial z_{i}}]. \eqno (5.11)
$$
By introducing the canonical coordinates:
$$
{1\over \sqrt{2}}(q_{i}+ip_{i})=\sqrt{M}{\nu_{i} \over \nu}\sin{\nu},
\eqno (5.12)
$$
one can show that (5.12) will be transformed into the following
canonical form:
$$
[ f,g ]_{PB}=\sum_{i,j=2}^3[{\partial f \over \partial q_{i}}
{\partial g \over \partial p_{j}}-{\partial f \over \partial
p_{j}}{\partial g \over \partial q_{i}}]. \eqno (5.13)
$$
The semiquantal dynamics of this many--body system with fixed
nucleons number M is determined by varying the effective action:
$$
S=\int [{1\over 2}({\vec p}\cdot d{\vec q}-{\vec q}\cdot d{\vec p})-
H({\vec p},{\vec q})dt]. \eqno (5.14)
$$
The result of such a variation is the set of classical--like
dynamical equations:
$$
{dq_{i} \over dt}={\partial H({\vec p},{\vec q}) \over \partial p_{i}}
, \quad \quad\quad
{dp_{i} \over dt}= -{\partial H({\vec p},{\vec q}) \over \partial q_{i}}
\eqno (5.15)
$$
with the Hamiltonian functions given by:
$$
H({\vec p},{\vec q})=<SU(3),{\hat \Phi}|{\hat H}|SU(3),{\hat \Phi}>.
\eqno (5.16)
$$
Equation (5.15) is in fact equivalent to the time--dependent
Hartree--Fock (TDHF) dynamical equations.
It should be noted that (5.15) is not a classical limit because
there is at least first--order quantum correlation in
$H({\vec p},{\vec q})$.

The phase space representation of the generators ${\hat G}_{ij}$ in the
coherent--states basis is:
$$
<|{\hat G}_{11}|>={1\over 2}(2M-{p}^2-{q}^2),
\eqno (5.17)
$$
$$
<|{\hat G}_{1i}|>={1\over 2}(q_{i}+ip_{i})
\sqrt{2M-{p}^2-{q}^2},
$$
$$
<|{\hat G}_{ij}|>={1\over 2}(q_{j}+ip_{j})
(q_{i}-ip_{i}),
$$
$$
<|{\hat G}_{ji}|>=<|{\hat G}_{ij}|>^{*},
$$
where $|>$ represents $|SU(3),{\hat \Phi}>$ and
$(q_{2}^2+q_{3}^2+p_{2}^2+p_{3}^2) \leq 2M$.

The quantum correlations of the quadratic function of the generators
can be computed and the results are:
$$
\Delta({\hat G}_{ij}^2)=<|{\hat G}_{ij}^2|>-<|{\hat G}_{ij}|>^2 =
-{1\over M}<|{\it {\hat G}}_{ij}|>^2 . \eqno (5.18)
$$
In this way the mean value of the hamiltonian in the coherent states
representation is:
$$
H({\vec p},{\vec q})={\epsilon_{1}\over 2}[2-(p_{2}^2+p_{3}^2
+q_{2}^2+q_{3}^2)]+{\epsilon_{2}\over 2}(p_{2}^2+q_{2}^2)+
{\epsilon_{3}\over 2}(p_{3}^2+q_{3}^2)+
 \eqno (5.19)
$$
$$
+{\chi\over 4}[1-{1\over M}]
[(p_{2}^2+p_{3}^2)^2-(q_{2}^2+q_{3}^2)^2
+(q_{2}^2-p_{2}^2)(q_{3}^2-p_{3}^2)+4(q_{2}q_{3}p_{2}p_{3})].
$$
It should be noted that the phase space has been scaled in (5.19)
such that $q_{2}^2+q_{3}^2+p_{2}^2+p_{3}^2 \leq 2$, with $\chi
=MV/\epsilon$.
The classical limit of the SU(3) model
can be realized by the limit $M \to \infty$, and, in this limit,
the quantum correlation
$\Delta({\it {\hat G}}_{ij}^2)$ goes to zero (see 5.18).
So the classical hamiltonian is ($\epsilon=1$):
$$
H_{cl} =
-1+{1 \over 2}q_{2}^2(1- \chi )+{1 \over 2}q_{3}^2(2- \chi )+{1 \over 2}
p_{2}^2(1+ \chi )+{1 \over 2}p_{3}^2(2+ \chi )+
$$
$$
+{1 \over 4} \chi [(q_{2}^2+q_{3}^2)^2-(p_{2}^2+p_{3}^2)^2
-(q_{2}^2-p_{2}^2)(q_{3}^2-p_{3}^2)-
4q_{2}q_{3}p_{2}p_{3}],  \eqno (5.20)
$$
with the phase space given by $\Omega=\{ (q_2,q_3,p_2,p_3)\in R^4:
(q_{2}^2+q_{3}^2+p_{2}^2+p_{3}^2) \leq 2\}$.
This hamiltonian represents two oscillators coupled
non--linearly by the parameter $\chi$ (the interaction between the
nucleons). The system is integrable if the interactions are
neglected but the interactions break this regular behaviour and produce
an order to chaos transition.

\vspace{0.5 cm}

{\bf 5.2 Classical calculations}
\vspace{0.5 cm}

Using the Hamilton equations of (5.20), in order to analyze the stability of
the system, we calculated the periodic orbits of this model:
$$
{\dot {\vec z}} = J \nabla H_{cl}({\vec z}, \chi ),  \eqno (5.21)
$$
where ${\vec z} = (q_{2},q_{3},p_{2},p_{3})$ , $\nabla =({\partial \over
\partial
q_{2}},{\partial \over \partial q_{3}},{\partial \over \partial p_{2}},{
\partial \over \partial p_{3}}),$ e $J$ is the 4x4 symplectic matrix:
$$
J = {\left(\matrix {0 & I \cr -I & 0 \cr } \right) } , \eqno (5.22)
$$
where $I$ is the 2x2 identity matrix.

To explore the phase space $\Omega$, we chose a 4--dimensional
lattice of initial conditions ${\vec z}_{0}={\vec z}(t=0)$,
${\vec z}_{0} \in\Omega$. The side of the lattice has about
$10^4$ initial conditions.

The time evolution of (5.21), with the initial condition ${\vec z}_{0}$, is
obtained using a $4^{th}$ order Runge--Kutta method [39].
The following function can then be constructed:
$$
d({\vec z}_{0},t)=d({\vec z}_{0},{\vec z}_{t})=|{\vec z}_{t}-{\vec z}_{0}|,
\eqno (5.23)
$$
where ${\vec z}_{t}={\vec z}(t)$. (5.23) is the function to minimize, with
minimum value equal to 0. To obtain a periodic orbit, it is
not actually necessary to vary all five parameters $({\vec z}_{0}, t)$;
in fact we can fix $E=H({\vec z}_{0})$ or t, which
correspond to focusing either on constant energy or constant
period in the E--T plot.

Another algorithm [40] has been used
to find the minimum of $d$. Since $d$ is the result of a complicated
calculation and its derivatives are not available, this algorithm is
particularly useful in our case because the derivatives of $d$ are not
required [41].

These periodic trajectories occur in one parameter families, each of
which describes a continuous curve in the energy--period plane (fig. 10);
just as in quantum mechanics the set of
energy levels is characteristic of a given hamiltonian $H$, so $H_{cl}$ can be
characterized by the plots E--T, where E is the energy of a periodic
trajectory and T the corresponding period.

If ${\vec w}$ is a vector of the tangent space $T\Omega_{{\vec z}}$
of the phase space manifold $\Omega$ at ${\vec z}$, its time evolution
is given by:
$$
{\dot{\vec w}}=J {\partial^2 H({\vec z})\over \partial {\vec z}^2}{\vec w}.
\eqno (5.24)
$$
By (5.21) and (5.24) the Lyapunov exponents can be calculated:
$$
\lambda ({\vec z})=\lim_{t\to\infty} {1\over t} \ln{|{\vec w}(t)|}.
\eqno (5.25)
$$
In terms of the stability matrix $M(0,t)$ [42], defined as:
$$
M_{ij}(0,t)={\partial z_{i}(t)\over \partial z_{j}(0)}  , \eqno (5.26)
$$
$\lambda ({\vec z})$ can be written:
$$
\lambda ({\vec z})=\lim_{t\to\infty}{1\over t}\ln{|M(0,t)|} , \eqno
(5.27)
$$
where $|M(0,t)|$ is a norm of the matrix $M(0,t)$.

This matrix can be calculated by solving its equations of motion:
$$
{\dot M}=J {\partial^2 H({\vec z})\over \partial {\vec z}^2}M,
\eqno (5.28a)
$$
with the initial conditions:
$$
M(0)=I \eqno (5.28b)
$$
where $I$ is the 4x4 identity matrix.
The calculation of the Lyapunov exponents is related to
that of the eigenvalues $\sigma_{i}$ of the matrix $M(0,T)$:
$$
\lambda_{i}({\vec z})={1\over T}\ln{\sigma_{i}}. \eqno (5.29)
$$

Now, using the unitariety of M, a periodic orbit is unstable if
$$
Tr(M)>4 \quad or \quad Tr(M)<0 \eqno (5.30a)
$$
and stable if
$$
0<Tr(M)<4. \eqno (5.31b)
$$

It is interesting to study the change of stability of periodic
trajectories as a function of $\chi$. In
figure 11 the ratio between the number of stable orbits and the number of
total orbits with period $T<30$ is plotted {\it vs} $\chi$.
As can be seen the sensitivity of the orbits to a small change of
$\chi$ is quite different
for different values of $\chi$, reflecting the transition order $\to$ chaos as
the coupling constant increases.

As mentioned in section 4.2, the calculation of the long--range behaviour
(non--universal) of $\Delta_{3}$ requires the knowledge of the shortest
period $T_{min}$ of the closed trajectories. For a {\it
chaotic} system with time--reversal symmetry [34]:
$$
\Delta_{3}^{\infty}={1\over \pi^2}\ln{(e L_{max})}-{1\over 8}, \eqno
(5.32a)
$$
where:
$$
L_{max}={\hbar \bar{\rho}\over T_{min}} \eqno (5.32b)
$$
and $\bar{\rho}$ is the average level density. Obviously $\bar{\rho}$ and
$T_{min}$ are functions of the coupling constant $\chi$.

Figure 12 shows $\Delta_{3}(L)$ for $\chi=100$, $T_{min}=0.04$,
$\bar{\rho}=115$. As can be seen, the breaking of universality is near L=38.
Obviously the semi--classical
estimate of the $\Delta_{3}(L)$ saturation does not agree with the
quantal calculations for $\chi =0.5$ and $\chi =2$ because the equation
(5.32) is valid only for a {\it chaotic} system [43].

Incidentally, $\AA$berg [44] gives a rough estimate of the shortest periodic
orbit in a nucleus:
$$
T_{min}\simeq {4R\over v_F}\simeq {4\cdot 1.2 A^{1\over 3}\over
0.3 c}, \eqno (5.33)
$$
and thus $L_{max}\simeq 80 A^{-{1\over 3}}\bar{\rho}$. For $^{152}Dy$ at
$E_{x}\simeq 3.0$ MeV, $\bar{\rho}\simeq 560$ MeV$^{-1}$, we get
$L_{max}\simeq 8000$, i.e. we would need a sequence of more than $8000$
energy levels with the same spin and parity. This example shows that we
should hardly expect to see a non--generic behaviour of long--range
spectrum fluctuations due to short periodic orbits in nuclei.

\vspace{0.5 cm}

{\bf 5.3 Quantum calculations}
\vspace{0.5 cm}

In order to find the eigenstates and eigenvalues of the hamiltonian (5.1), we
need to find a complete set of basis states. A natural basis can be written
$|bc>$ meaning b particles in the second level, c in the third and, of
course, $M-b-c$ in the first level; in this way $|00>$ is the ground state
with all the particles in the lowest level [36,37]. We can write the general
basis state:
$$
|bc>=\sqrt{1\over b!c!}{\hat G}_{21}^{b}{\hat G}_{31}^{c}|00>, \eqno (5.34)
$$
with $\sqrt{1\over b!c!}$ the normalizing constant.

{}From the commutation relation (5.4) we can calculate the expectation values
of
${\hat H}\over M$ and thus, eigenvalues and eigenstates of
${\hat H}\over M$;
in this way the energy spectrum range is independent of the
number of the particles:
$$
<b^{'}c^{'}|{{\hat H}\over M}|bc>={1\over M}(-M+b+2c)\delta_{bb^{'}}
\delta_{cc^{'}}-{\chi \over 2M^{2}}Q_{b{'}c^{'},bc},
\eqno (5.35)
$$
where:
$$
Q_{b{'}c^{'},bc}=\sqrt{b(b-1)(M-b-c+1)(M-b-c+2)}\delta_{b-2,b^{'}}
\delta_{cc^{'}}
\eqno (5.36)
$$
$$
+\sqrt{(b+1)(b+2)(M-b-c)(M-b-c-1)}\delta_{b+2,b^{'}}\delta_{cc^{'}}
$$
$$
+\sqrt{c(c-1)(M-b-c+1)(M-b-c+2)}\delta_{b,b^{'}}\delta_{c-2,c^{'}}
$$
$$
+\sqrt{(c+1)(c+2)(M-b-c)(M-b-c-1)}\delta_{b,b^{'}}\delta_{c+2,c^{'}}
$$
$$
+\sqrt{(b+1)(b+2)c(c-1)}\delta_{b+2,b^{'}}\delta_{c-2,c^{'}}
$$
$$
+\sqrt{b(b-1)(c+1)(c+2)}\delta_{b-2,b^{'}}\delta_{c+2,c^{'}}
$$
and $\chi =MV/\epsilon$.
The expectation value $<{{\hat H}\over M}>$ is real and
symmetric. For any given number
of particles M, we can set up the complete basis state, write down the matrix
elements of $<{{\hat H}\over M}>$ and then diagonalize $<{H\over M}>$
to find its
eigenvalues. $<{H\over M}>$ connects only states with $\Delta b=-2,0,2$ and
$\Delta c=-2,0,2$ which makes the problem easier. We group states with b,c
even; b,c odd; b even and c odd; b odd and c even. This means that
$<{{\hat H}\over M}>$ becomes block diagonal containing 4 blocks which can be
diagonalized separately; these matrices are referred to as ee, oo, oe and eo.

To separate regular and chaotic state in a quantum system a powerful
tool is the study of spectral statistics. We obtain [45] a good
agreement with GOE in the classically
chaotic region and, in the classically regular region, a good agreement with
Poisson statistic (see fig. 13). In this figure the continuous curve is
the Brody distribution, discussed in section 6.5.

On the basis of the semiclassical torus
quantization, the presence of crossings in a small $\chi$ neighbourhood may
be interpreted as the signature of quasi crossing in the true system, and
thus as a signature of torus destruction, when the exact levels are ``split"
at the crossing [46--48].

Another method to study the quantum stochasticity is
the sensibility of the energy levels to variations of the perturbation
parameter; the behaviour of the curvature of the energy level $E(\chi )$
in a small $\chi$ neighbourhood is an example [49].

When the parameter $\chi =0$, then the hamiltonian represents two oscillators,
and there are many degenerations, but for $\chi \neq 0$ these degenerations
are broken.
We calculated [45], for a large number of particles (semiclassical limit)
the density of quasi crossings outside the degeneration region as function of
the parameter $\chi$:
$$
\rho (\chi )={\Delta N\over \Delta\chi}, \eqno (5.37)
$$
where $\Delta N$ is the number of quasi crossings in the parameter range
$\Delta \chi$. To obtain $\Delta N$, we fixed three values $\chi -
\Delta \chi$, $\chi$ and $\chi +\Delta \chi$ and imposed that:
$$
s_{i}(\chi -\Delta \chi ) > s_{i}(\chi) \eqno (5.38a)
$$
$$
s_{i}(\chi +\Delta \chi ) > s_{i}(\chi) \eqno (5.38b)
$$
where $s_{i}(\chi )=E_{i+1}(\chi ) - E_{i}(\chi )$.
The results (fig. 14) show for all the classes a maximum of quasi
crossings for $\chi =2$ in agreement with the transition to chaos of
fig. 12.

In order to study the sensitivity of the energy levels to small changes of the
parameter $\chi$, we have used the statistic $\Delta^2 (E)$ [50],
defined as:
$$
\Delta^{2}(E_{i})=|{E_{i}(\chi +\Delta \chi )+E_{i}(\chi -\Delta \chi )-
2E_{i}(\chi )}|. \eqno (5.39)
$$
This statistic measures the curvature of $E_{i}$ in a small range
$\Delta \chi$. Fig. 15 shows $\Delta^2 (E)$ in the [--0.5,0.5] energy
range for different values of $\chi$; in this case we note the
formation of a peak in correspondence to the $\chi =2$ region.

\newpage

\par
{\bf 6. Wave function behaviour and EM decay of regular and chaotic
states}
\vskip 0.5 truecm
\par
As we have seen in the previous sections, it is now well known that in
atomic nuclei there are two different states: regular and chaotic. These
states can be highlighted with the aid of various statistics.
The statistics we have considered before concern the
eigenvalues of the system, and the most frequently used are those of level
spacing; i.e. the nearest neighbour distribution, $P(s)$, and the
stiffness of the spectrum $\Delta_3(L)$.
Since these statistics display clearly different
behaviour depending on the regularity or chaoticity of the nucleus, they
are very useful to distinguish between the two regimes and, at the same
time, characterize the transition from one region to the other.

However, it is well known that, in general, the wave functions and
the transition probabilities are much more sensitive to the purity of
states than the eigenvalues are.
Consequently, it is interesting to undertake the study of statistics
directly related to the wave functions with the aim of characterizing the
ordered or chaotic behaviour of atomic nuclei.

Using the SU(3) schematic shell model for the nucleus, described in Section 5,
we have studied the various statistics concerning the
wave functions, such as the intensity of the momenta of the wave function
$I_m$, the information measure or entropy $S$, and
the correlation functions $K_{n,m}$.

The momenta $I_m$ and the information measure $S$ are defined [51]:
$$
I_{m}=\sum_{b,c}|\psi (b,c)|^{2m}, \eqno (6.1)
$$
$$
S=-\sum_{b,c}|\psi (b,c)|^2\ln{|\psi (b,c)|^2}. \eqno (6.2)
$$
Figure 16 shows that the lower momenta (m=2,3,4) diverge in the
regular region but have a constant behaviour in the chaotic region,
as predicted by [52].

The information measure $S$ also shows different behaviours in the two regions,
assuming high values in the chaotic region and low values in the regular region
because wave functions are localized (see fig. 17).

Another way to separate regular from chaotic states is with the help of
the transition probabilities among different states [53]. Only a few
strong transitions between regular states and many weak transitions between
chaotic states are expected. The transition probability is defined:
$$
P_{ij}^{(k)}=|<\psi_{i}^{\tau}|T_{k}|\psi_{j}^{\tau^{'}}>|^2, \eqno (6.3)
$$
where:
$$
T_{1}={1\over 2}(G_{12}-G_{21}),\quad T_{2}={1\over 2}(G_{13}-G_{31}),\quad
T_{3}={1\over 2}(G_{23}-G_{32}). \eqno (6.4)
$$
$$
|\psi_{i}^{\tau}>=\sum_{\alpha =1}^{d_{\tau}}x_{i,\alpha}^{\tau}|b,c> \eqno
(6.5)
$$
is the wave function of the j--state of class $\tau$ with dimension $d_{\tau}$
(ee, oo, eo and oe). Scaling  $P_{ij}^{k}$ with a local mean value
$<P_{ij}^{k}>$, the distributions $P_{ij}^{k}\over <P_{ij}^{k}>$ follow
the above previsions very well (see fig. 18).

The previous results clearly show that the wave function behaviour of
atomic nuclei is a very powerful tool for distinguishing regular and
chaotic regimes. In particular, the study of the transition
probabilities between different states is of great interest.

\vskip 0.5 truecm
{\bf 6.1 Decay of regular and chaotic states}
\vskip 0.5 truecm
\par
Recently, new interest has been shown in the study of the decay
of unstable quantum states. In this section we briefly review the
literature of major interest.
The section is organized as follows: first, we look at the current
experimental stage to illustrate the reasons and interest for
studying the decay of quantum systems. Secondly, we review the main
current theoretical approaches to the problem. We begin with the study
of "warm" rotational nuclei,
to continue with that of unstable systems described by means of an
anti-hermitian effective hamiltonian. Finally, we briefly discuss
the Interacting Boson Model (IBM) that affords a wide spectrum of
applications, from vibrational to rotational nuclei as well as
transitional regions.

\vskip 0.5 truecm
{\bf 6.2 Experimental stage}
\vskip 0.5 truecm
\par
The new generation of experimental instruments like GASP,
EUROBALL, EUROGAM, etc [54], which provide a resolution in the order of
a hundred times greater than that of the previous generation,
will open a new way in nuclear spectroscopy research.
Over the last few years, as discussed above, great interest has been shown in
the study of the properties of the atomic nuclei from a statistical
point of view. This has made it possible, with the aid of the
Random Matrix Theory [14], to distinguish between regular
and chaotic regimes, the onset of chaos, the transition from
one regime to the other, etc.

Let us now turn our attention to the recent statistical
analysis of $\gamma$-ray spectra of "warm" rotational nuclei
by Garrett, Hagemann and Herskind [55]. In their work,
two and three-fold energy correlation measurements between
gamma rays were analyzed to investigate high spin states
of the so called quasi-continuum that spreads up to 5-10 MeV
above the yrast line. This study had already been proposed by
Guhr and Weidenm\"uller [56] who stressed the coexistence
of collectivity and chaos in nuclei at several MeV of excitation energy;
they suggested the study of stretched collective E2 decays of high
spin states in deformed even-even nuclei at, or above, the yrast line.
It was shown that the collective E2 transitions from states of
spin $I$, located several MeV above the yrast line, populate many
states (of the same parity) of $(I-2)$ spin rather than a single one
(as happens with the E2 transitions from states near the yrast line,
where a well defined rotational band structure exists). Thus the transition
strength is spread over an energy interval that is called rotational
damping width.
The analysis of the group of Herskind is indeed oriented
in this direction. By performing a fluctuation analysis of the
2--D $E_{\gamma_1}$x$E_{\gamma_2}$ spectrum they obtained information
on the number of decay routes the nuclear decay flow takes to go
from the initial high spin state to the damped region (which is
supposed to be in the diagonal valley, $x=y$, of the 2--D spectra).
A schematic illustration of the average flow of $\gamma$-ray
decay from high-spin states induced by heavy-ion compound reactions
is shown in fig. 19.

The 2--D and 3--D fluctuation analysis method developed by this
group [55] has been applied to the spectra
from a triple coincidence experiment
made by the Manchester--Daresbury--Copenhagen--Bonn collaboration at the
Daresbury Tandem Accelerator. The reaction consisted in bombarding
$^{124}Sn$ by a beam of $^{48}Ca$ of $215$ MeV, to form the residual
nuclei of $^{168}Yb$ at the highest spin $I_{max} = 60 \hbar$.
{}From the fluctuation analysis done on $^{168}Yb$ they concluded that
the rotational correlations originating from well defined rotational
band structures only exist up to 1 MeV above the yrast line.
Over this limit the main decay routes spread out into many branches.
This may be accounted
for by the rotational damping phenomenon, as Guhr and Weidenm\"uller
[56] had previously suggested in 1989 from a theoretical calculation
with a GOE hamiltonian weakly perturbed by a residual interaction.
In the conclusions of this work, the authors proposed the measurement
of the spreading width of
E2--strength for a high--spin state well above the yrast line,
to determine the possible onset of chaos in "warm" deformed
nuclei. The fluctuation analysis mentioned above has in fact taken up
this suggestion.

The first conclusions extracted from this new analysis of $\gamma$-ray data
suggest that there is much interest to be had from the study of states
near and above the yrast line, "warm" rotational nuclei and
the decay of unstable quantum states.
New theoretical attention has been devoted to these phenomena.

\vskip 0.5 truecm
{\bf 6.3 Rotational damping motion}
\vskip 0.5 truecm

Due to the recent experiments in $\gamma$-ray detection,
a renewed interest has developed in the the study of the
decay of chaotic states and the possibility of clarifying
the mechanisms by which an
excited nucleus ($6$--$8$ MeV over yrast line) undergoes a transition
from a regular to a chaotic behaviour.

In this sense, the work of Matsuo, Dossing, Herskind and Frauendorf [57]
is very interesting. They investigate the properties of energy levels and
rotational transitions as a function of the excitation energy,
in "warm" deformed nuclei, and show that,
at an excitation energy of the order of $E_x \simeq 8$ MeV,
there is considerable fragmentation of the rotational E2 strength
distribution when it is represented against the $\gamma$ energy
(see fig. 20). This fragmentation with increasing energy
is due to the mixing of the rotational bands caused by
the Surface Delta Interaction (SDI):
$$
V(1,2) = 4 \pi V_0 \delta (r_1 - R_0) \delta (r_2 -R_0)
\cdot \sum_{\lambda,\mu} Y_{\lambda,\mu}^*(r_1) \cdot
Y_{\lambda,\mu}(r_2),
\eqno (6.6)
$$
which contains all the possible multipolarities. Actually, the
high multipole components of this SDI are shown to be
responsible for the mixing of rotational bands which is reflected
in fluctuations typical of chaotic behaviour.
The strength fluctuations of the E2 transitions, for
instance, obey a Porter--Thomas distribution above a certain excitation
energy, assumed to be the threshold for the realization
of quantum chaos in the system (fig. 21).

In particular, the same authors have also studied high spin levels
in nuclei with a cranked shell model extended
to include residual two-body interactions [58].
Here again the residual interaction
induces the transition from a regular to a chaotic regime,
since when the rotational bands do not interact, $\gamma$-ray energies
behave like random variables, i.e., they obey a Poisson
distribution typical of a regular regime. On the contrary,
when the residual interaction is added,
at an excitation energy over $600$ KeV, a gradual rotational
damping is established, and, at $1.8$ MeV above the yrast line,
the complete damping is observed and typical GOE fluctuations
of the energy levels and transition strengths are produced (fig. 22).

A similar approach is used by $\AA$berg [59] to study
the rotational damping of rapidly rotating
nuclei. As in the preceeding approach, a cranked Nilsson model
with a schematic residual interaction
is used to study the distribution of the E2-strength, and
$\gamma-\gamma$ correlations.
In particular the normal deformed (ND) $^{168}Yb$
and the superdeformed (SD) $^{152}Dy$ are studied, and the different
behaviours of these nuclei with increasing energy
are compared.
The main difference between the approach of $\AA$berg and that
of the authors of ref. [58] is that in the former the energies of the
np-nh states are given in the laboratory frame as functions of
the angular momentum I. So, diagonalization is performed
for each fixed I value. This is reflected in a change of the
individual energies, but the conclusions regarding the
different statistical properties are the same.
The $\AA$berg approach is perhaps more useful if a direct comparison
with $E_\gamma-E_\gamma$ experimental spectra is to be carried out.
The "classical" eigenvalue statistics are
studied (i.e. $P(s)$, $\Delta_3(L)$) with special regard to
the $\Delta_3(L)$ since it gives a measure of long-range
correlations. In this model the transition from order to
chaos is brought about by changing the strength of the two-body
force ($\Delta$) which is taken as a parameter, $V_2=\pm \Delta$,
where the sign is chosen randomly to avoid coherent effects.
Following the procedure of Brody [60], the $\Delta_3(L)$ spectrum
is fitted with a single parameter, $q$, to study the mixing
between the Poisson and GOE distribution:
$$
\Delta_3 (L,q)=\Delta_3^P [(1-q)L] + \Delta_3^{GOE} (qL).
\eqno (6.7)
$$
The dynamical properties are studied in this model and again
a rotational damping is manifested in the fragmentation of E2
strength in many daughter states. The average standard deviation
of the E2-strength function saturates for sufficiently large
values of the strength of the two-body residual interaction,
following the same trend as $q$ parameter, which
gives the mixing of a Poisson and GOE behaviour of spectrum properties
(see figs. 23--24).

In addition to the previously mentioned rotational damping,
$\AA$berg also studies motional narrowing, i.e., the decrease of
the width of nuclear magnetic resonances when the temperature
increases, a phenomenon which may be studied in a simple two-band
model. The narrowing of the strength function is accomplished
by a change from a Gaussian shaped strength to a Breit-Wigner
shape. This phenomenon is understood in terms of time scales:
when the time scale of the fluctuations in available rotational
frequencies is much less than the time the wave function takes
to spread out over the basis states, the intensity spectrum
is Gaussian, while in the opposite case the corresponding spectrum
is of the Breit-Wigner type.

Since, due to the motional narrowing the E2 strength may
become very narrow at the end, a small number of relatively
strong E2 transitions may be observed, and, consequently,
a "suppression of chaos". In fact the transition entropy defined
as [44]:
$$
H^{T,\alpha} = - \sum_{\alpha'}  [M_{\alpha \alpha'}]^2 \cdot
ln ( [M_{\alpha \alpha'}]^2),
\eqno (6.8)
$$
where:
$$
M_{\alpha \alpha'} (I)=<\alpha',I-2|M(E2)|\alpha,I>
\eqno (6.9)
$$
decreases when motional narrowing sets in.

The $\gamma-\gamma$ correlations between consecutive
$\gamma$-rays are also analyzed to study the fragmentation of
E2 strength. Correlations are revealed at small excitation energies or
when a small $\Delta$ is used, while
at high excitation energies these correlations disappear.

\vskip 0.5 truecm
{\bf 6.4 The study of unstable systems using a complex effective
hamiltonian}
\vskip 0.5 truecm

A different and more general approach to the study of unstable
quantum systems is that of Sokolov and Zelevinsky [61] and
more recently that of Mizutori and Zelevinsky [62], in which the
statistical theory of spectra, formulated in terms of random
matrix theory, is generalized to treat unstable states,
i.e. those coupled to open channels. In
this way the influence of the coupling with the continuum
on the properties of internal states can be better understood.

In general, as we have seen up to now, the level statistics are treated as
if the states were stationary but, in reality, all
excited states are resonances embedded in the continuum.
So if one intends to study the most general states
of a system (including the unstable ones) a new approach
to the problem has to be used.

When the widths of the states are small as compared with level spacings,
the approximation of discrete levels is reasonable for long-lived
states. However, when the widths increase and the levels overlap,
the effects of the finite lifetime of those become crucial
and the application of a random matrix ensemble of
hermitian hamiltonians is not enough appropriate.

Since the standard gaussian
ensembles are applicable to discrete stationary states, while
an excited state decays via open
channels, the use of a non-hermitian effective hamiltonian
H is imposed to take into account the width of the states
due to their finite life [61,62].

Within this model the reaction amplitudes are represented, using
the general theory of resonant nuclear reactions,
as sums of pole terms in the complex plane. Such poles,
$\epsilon_n = E_n -(1/2)i \Gamma_n$ are the
eigenvalues of the non-hermitian effective hamiltonian
$\bar{H} = H - (i/2) W$ where $H$ is the hermitian part belonging
to the GOE. The amplitudes of the antihermitian part, $W$, have
a separable structure $W_{nm} = \Sigma_c A_n^c A_m^c$ due to the
unitarity of the scattering matrix, where $A_n^c$ are the amplitudes
for the decay of intrinsic states of the system, $|n>$ ($n$=1,...N) into
different channels, $c$ ($c$=1,....k) and that must be real according to
time reversal invariance of the full hamiltonian, $\bar{H}$.
Those eigenvalues correspond to the intermediate unstable states of energies
$E_n$ and widths $\Gamma_n$.

Decaying systems are hence described by ensembles of
random non-hermitian hamiltonians represented by N-dimensional
matrices where the hermitian part of the effective hamiltonian, H,
belongs to the GOE. The decay amplitudes $A_n^c$, on the other hand,
are assumed to be Gaussian random variables completely uncorrelated with
the matrix elements of $H$ and among each other.

Within this model the distribution function of complex
energies, the level spacing distribution and the width
distribution are studied in ref. [61,62] for weak and strong coupling
to the continuum, as well as for the transitional region.

The authors of ref. [62] discussed in great detail the single channel
case: $k=1$. They assumed that $<A_n>=0$, $<H_{mn}>=0$, and:
$$
<H_{nn^{'}}H_{mm^{'}}>={a^2\over
4N}(\delta_{mn^{'}}\delta_{m^{'}n}+\delta_{mn}\delta_{m^{'}n^{'}}).
\eqno (6.10)
$$
The strength of coupling to the continuum
is given by the overlap parameter $\chi = \eta /a = 2<\Gamma> / D$
where $\eta = \sum_j <\Gamma_j>$; $D$ is the mean level spacing.

When the overlap parameter is small ($\chi << 1$), that is, when coupling
with the continuum is weak, the hermitian part H dominates, while the
antihermitian part, $W$, prevails in the opposite limit ($\chi >> 1$).

Typical signatures of the transition from weak to strong
coupling are observed: first, concerning the level spacing
distribution, which gives the simplest characterization
of spectral correlations, it is observed that for
a weak coupling (where the hermitian part, $H$, of the
effective hamiltonian dominates) the distribution of levels
corresponds to a Wigner distribution, and deviations from it
are small. As the coupling with the continuum increases, the Wigner
function ceases to be a good approximation to the GOE and a new
feature is observed: the level repulsion at short distances
disappears.

The authors of ref. [62] have fitted the $P(s)$
distribution by a simple one-parameter superposition of
the normalized Wigner, $P_W$, and Gaussian, $P_G$ distributions,
that is:
$$
P(s) = \alpha P_W(s) + (1 - \alpha) P_G(s)
={1\over \Delta}[{\alpha\over \Delta} s + (1 - \alpha)\sqrt{2\over \pi}]
\exp{(- {s^2 \over 2\Delta^2})},
\eqno (6.11)
$$
where the parameter $\Delta$ is determined, for a fixed $\alpha$,
by the mean level spacing:
$$
D = \Delta [ \sqrt{\pi\over 2} \alpha +\sqrt{2\over \pi}(1 - \alpha)].
\eqno (6.12)
$$
In the fig. 25 the trend of the mixing parameter $\alpha$ for
the $P(s)$ as a function of the coupling constant $\chi$ is plotted,
showing a pronounced minimum ($\alpha =0.74$) at the transition point
$\chi =1$. Both small and large values of $\chi$ correspond
to the nearest level statistics close to a Wigner distribution $P_W$.
It is well known that the
GOE implies rigid spectra with small level spacing fluctuations. So, we
can conclude that coupling to the decay channels
softens the spacing distribution so that the spacing
fluctuations increase gradually in the transition region.

As regards the distribution of complex energies, it is
has been observed that, for a random matrix ensemble,
the widths distribution corresponds to a Porter-Thomas
one, i.e.:
$$
P^{PT} (\Gamma) = {1\over \sqrt{2 \pi <\Gamma> /\Gamma)}}
\exp{(-{<\Gamma>\over 2}\Gamma)}.
\eqno (6.13)
$$
This, however, is no longer correct when the level spacings become
very small compared to the widths, i.e., when the
overlap between neighbour states is considerable. This
is a consequence of the coupling to the continuum, since for
weak coupling the typical widths are small compared to
spacings. When, however, the coupling to the continuum is important,
the widths become larger than the mean level spacing. This
behaviour had already been observed for complex random matrix
ensembles [63] and in chaotic dissipative systems [64].

The most interesting statistic for highlighting
the phase transition is the width distribution. In fig. 26
the widths, plotted as a function of the coupling parameter $\chi$,
show a collectivization. The mean fraction $< \Gamma_1 /\sum_j \Gamma_j>$
of the total width is accumulated by the broadest state $\Gamma_1$.

In conclusion, we can say that within these works
a standard statistical spectroscopy
of discrete levels and unstable states, as well as for the
transitional region, has been performed using ensembles of random
non-hermitian hamiltonians represented by N-dimensional
matrices. They show [61,62] that the instability of states changes
the statistics remarkably, removing level repulsion when
distances are smaller than widths.
When the coupling to the continuum increases, that is,
when the matrix elements of the antihermitian part of the
effective hamiltonian become comparable with the spacings
of eigenvalues of the hermitian part, a transition to a new
regime occurs.

Related to the study of unstable systems and in the
same spirit as the preceeding work, a paper
by Haake et al. [65] is noteworthy. The authors study the level density of
different classes of random non-hermitian
matrices, $\bar{H} = H + i \Gamma$, where the damping, $\Gamma$, is chosen
quadratic in Gaussian random numbers, to describe the decay
of resonances through various channels.
When the notion of level spacing is extended to the Euclidian
distance between complex eigenvalues of non-hermitian operators,
two different behaviours of $P(s)$ statistic (as in the ordinary
case of real eigenvalues with the Wigner and Poisson statistics)
allow the distinction between regular and chaotic dynamics: cubic
repulsion tends to be typical under conditions of global classical
chaos while linear repulsion signals classical integrability. This
classification also arises for dissipative systems [66].

\vskip 0.5 truecm
{\bf 6.5 The IBM model in the study of unstable systems}
\vskip 0.5 truecm

The profound understanding of the statistics of low-lying
levels of nuclei and the underlying chaotic dynamics requires
realistic theoretical models of the nucleus. In general, the
models used have only two degrees of freedom.
However, since the quadrupole deformation plays an important
role in collective nuclear dynamics, a realistic model
requires at least five degrees of freedom.

The work of Whelan and Alhassid [67] have been along these lines.
Using the Interacting Boson Model (IBM) [3], whose classical limit
is obtained with coherent states and which provides a good
description of low-lying energy levels and EM transitions of
heavy nuclei, they studied, classically and quantum mechanically,
the chaotic properties of low--lying collective states of atomic nuclei.
The six degrees of freedom come from
one monopole s-boson ($0^+$) and five
$d_{\mu}$ $(\mu=-2,...,2)$ bosons $(2^+)$ with which a U(6) dynamical
algebra is constructed. The most general hamiltonian is then
constructed with all one and two-body scalars that conserve the
total number of bosons, $N= ss^+ +\sum_{\mu} d_{\mu}^+d_{\mu}$.

The most useful parameterization of the IBM hamiltonian is given by:
$$
H = E_0 + c_0 n_d + c_2 Q^{\chi} \cdot Q^{\chi} + c_1 L^2,
\eqno (6.14)
$$
where $n_d = d^+ \cdot \bar{d}$ is the number of d-bosons,
$L$ is the angular momentum and $Q^{\chi}$ is the quadrupole operator:
$$
Q^{\chi} = (d^+ \times \bar{s} + s^+ \times \bar{d} )^{(2)}
+\chi (d^+ \times \bar{d})^{(2)},
\eqno(6.15)
$$
that depends on a parameter $\chi$, and where
$\bar{d}_{\mu} = (-)^{\mu} d_{\mu}$
so that $\bar{d}_{\mu}$ transforms under rotations like $d_{\mu}^+$.

Its classical limit is obtained for the number of bosons
going to infinity, since $1/N$ plays the role of $\hbar$.
That classical limit is:
$$
h = \epsilon_0 + \bar{c} [ \eta n_d -(1 - \eta) q^{\chi} \cdot q^{\chi}]
+c_1 L^2,
\eqno (6.16)
$$
where:
$$
{\eta \over 1 - \eta } = - {c_0 \over N c_2},\;\;\;
\bar{c} = {c_0 \over \eta},
\eqno (6.17)
$$
with $0< \eta < 1$.
In an algebraic model like this, a dynamical symmetry occurs
when the hamiltonian can be written as a function of the Casimir
invariants C of a chain of subalgebras of the original algebra.
So, the authors of ref. [67]
analyze the rotational nuclei (SU(3) symmetry),
vibrational nuclei (U(5) symmetry) and $\gamma$-unstable nuclei
(O(6) symmetry):

U(6)$\supset$U(5)$\supset$O(5)$\supset$O(3) (vibrational nuclei)

U(6)$\supset$SU(3)$\supset$O(3) (rotational nuclei)

U(6)$\supset$O(6)$\supset$O(5)$\supset$O(3) ($\gamma$-unstable nuclei).

The authors study the character of the
classical dynamics of a nucleus described by the classical
limit of the quantum IBM hamiltonian.
As a first result, they found that the quantal fluctuations,
which are well correlated with the classical results, are independent
of the number of bosons. The transitions between the
different dynamical symmetries of the model (i.e., between the
different dynamics of nuclei) with the variation of the parameter
$\chi$, of both classical and quantal hamiltonian, are studied.

The transition between deformed rotational nuclei (SU(3)) and spherical
vibrational nuclei (U(5)) is obtained for $\chi = -1/2 \sqrt{7}$ and
$0<\eta <1$. With these parameters, classical
and quantal measures of chaos are studied:
the average maximal Lyapunov exponent $\lambda$ and the fraction of chaotic
trajectories $\sigma$, on the one hand, and the parameter $\omega$ of the
Brody distribution of level spacing [60] on the other:
$$
P_{\omega} (s) = A s^{\omega} \exp{(- \alpha s^{\omega +1})}.
\eqno (6.18)
$$
The Brody distribution, as the $\Delta_3^q(L)$ mentioned above,
interpolates between a Poisson distribution ($\omega =0$) for a regular
system and a Wigner one ($\omega = 1$) for a chaotic system. $\alpha$
and $A$ are chosen such that $P_{\omega}(s)$ is normalized to $1$
and $<S> = 1$. A second quantal measure of chaos is $\nu$, which characterizes
the $B(E2: I\to I)$ distribution for the levels with $I = 2^+$:
$$
P_{\nu} (y) = {1\over \Gamma({\nu \over 2})}
({\nu \over 2 <y>})^{\nu \over 2}
y^{{\nu \over 2} -1} \exp{(- {\nu y \over 2 <y>})},
\eqno (6.19)
$$
where:
$$
y = |< f | T | i > |^2,
\eqno (6.20)
$$
$|i>$ and $|f>$ being, respectively, the initial and final states and
$T$ the transition operator. $P(y)dy$ is the probability of having
intensity $P(y)$ in the interval $dy$ around $y$.
$P(y)$ reduces to a Porter-Thomas distribution for $\nu =1$ (chaotic
limit) and, as the system becomes more regular, $\nu$ decreases towards small
positive values.

The maximum of chaos is obtained for $\eta = 0.5-0.7$.
The quantal results show a
strong correlation with the classical dynamics, since $\omega$ and $\nu$
are largest around $\eta = 0.5-0.7$.
Secondly, the transition between rotational nuclei (SU(3)) and
$\gamma$-unstable
nuclei (O(6)) is obtained for $\eta = 0$ and $ -1/{2} \sqrt{7}<\chi <0$,
chaos being settled for intermediate values of $\chi$.
Finally, the transition between $\gamma$-unstable (O(6)) and vibrational
nuclei (U(5)), for $\chi =0$ and $0<\eta <1$, is always completely
regular.

The most important result is, on the one hand, the discovery of
a new nearly regular region that lies between a rotational (SU(3))
and vibrational (U(5)) regime of the nucleus, and that is not
related to any of the known dynamical symmetries of the model.
Since the fraction of chaotic trajectories
$(\sigma )$ is $<0.3$, that region is not completely regular.
The signatures of this nearly regular region are identified by
the rather sharp minimum in all measures of chaos, as seen in fig. 27.
The authors suggest
that this new nearly regular region may be connected with a previously
unknown approximate symmetry of the model.
On the other hand, a spin dependence of the degree of chaoticity is
revealed. When both classical ($\lambda$,$\sigma$) and quantal
($\omega$,$\nu$,$q$) measures of chaos are plotted versus spin,
an interesting dependence is found:
in all measures (except $\lambda$) a weak dependence
on the spin, at low and medium spins, is found. However,
at high spins ($I > 20\hbar$)
there is a rapid decrease of chaoticity and the motion becomes regular.

In conclusion, strong correlations are observed between the onset
of classical and "quantum" chaos. Although the IBM model can be useful
for the study of the degree of chaoticity of low-lying collective
states of nuclei, at higher spin and/or energy, bosons may break into
quasi-particles and it is also important to take into account
the additional fermionic degrees of freedom,
to obtain a realistic description [4].

In this section we have presented some of the most
interesting recent works, which have attempted to explain
the mechanisms of the decay of quantum states and its connection with the
onset of chaos.

The theoretical efforts are also supported by the new detector
generation, which, with their higher resolution, can provide
a new and profound knowledge in the field of nuclear spectroscopy.

\newpage
\par
{\bf 7. Conclusions and open problems}
\vskip 0.5 truecm

To conclude this presentation of some regular and chaotic aspects in nuclear
dynamics, it is important to remember two points. The first is
the good agreement between the theoretical previsions concerning the
Poisson$\to$GOE transition and the experimental data;
the second is the possibility of studying
the order to chaos transition using simple models.

As pointed out by many authors (see for instance ref. [1]),
the static nuclear mean field is too regular (as witnessed
by the existence of shell structure) to be responsible of chaotic behaviour,
and chaos must be caused by residual interaction, as shown in the
models discussed above.
Because of the exclusion principle, the role of the latter increases
with the excitation energy. This makes one wonder whether nuclear motion
might be more regular in the ground--state region than, say, at neutron
threshold. Indeed, at several MeV excitation energy, collective states and
giant resonances acquire substantial spreading widths [1]. They lose their
individuality as excitation states, and show up only as bumps in strength
functions. The same kind of statement applies also to rotational bands 1 or
2 MeV above the yrast line. Much work remains to be done to elucidate the
role of chaotic motion and its interplay with collectivity.

There is also the problem of showing which properties of the nucleon--nucleon
interaction justify the use of random matrix models. These, according to
ref. [1], are related to nonintegrability and chaos, but the matter
requires further research to obtain a deeper understanding of the problem.

We have shown that chaos in nuclei is a rich field; it permeates many
aspects of nuclear structure and reactions, and gives rise to new
analyses of the experimental data.

\newpage

{\large \bf Figure Captions}
\vspace{0.6 cm}

{\bf Figure 1}: The total neutron cross--section on $^{232}Th$
{\it vs} neutron energy (adapted from [7]).

{\bf Figure 2}: The NDE experimental results compared with the
theoretical predictions (adapted from [19]).

{\bf Figure 3}: $P(s)$ for "cold" deformed rare--earth nuclei
(adapted form [20]).

{\bf Figure 4}: Spectral statistics for nuclei with $24<A<244$ and excitation
energy of few MeV; (a) $2^+$ and $4^+$ states, even--even nuclei;
(b) all other states, even--even nuclei;
(c) states with non--natural parity, odd--odd nuclei;
(d) states with natural parity, odd--odd nuclei (adapted from [21]).

{\bf Figure 5}: A diagramatic representation
of the ranges of excitation energy
$E_x$ and angular momentum $I$ associated with each set of data
previously discussed (adapted form [20]).

{\bf Figure 6}: Closed (a) and not closed (b) orbits on the torus.

{\bf Figure 7}: A Poincar\`e section.

{\bf Figure 8}: Poincar\`e sections: (a) regular closed orbit, (b)
regular not closed orbit, (c) chaotic orbit.

{\bf Figure 9}: $P(s)$ and $\Delta_3(L)$ for billiards. Regular billiard:
circular boundary (left), chaotic billiard: stadium boundary corresponding
to eigenfunctions with odd--odd symmetry (right) (adapted from [29]).

{\bf Figure 10}: $E-T$ plots for different values of $\chi$ and for different
families: $\triangle$ initial conditions along the axis $q_1$,
$\Box$ initial conditions along the axis $q_2$,
$\bigcirc$ initial conditions near the minima of
the static potential: (a) $\chi =0.5$, (b) $\chi =2$, (c) $\chi =100$
(adapted from [43]).

{\bf Figure 11}: Ratio between the number of stable periodic orbits and
the number of total periodic orbits {\it vs} $\chi$ (adapted from [45]).

{\bf Figure 12}: Universal and long--range behaviour of $\Delta_3(L)$ {\it vs}
$L$. The continuous curve is the GOE behaviour (adapted from [43]).

{\bf Figure 13}: $P(s)$ {\it vs} $s$ for the $eo$ class with $M=102$; (a)
$\chi =0.75$, (b) $\chi =2$, (c) $\chi =3$. The continuous curve is the
Brody distribution (6.12) with $\omega = 1-{\Delta N_{st}/\Delta
N_{tot}}$ (adapted from [45]).

{\bf Figure 14}: Density of quasi--crossing {\it vs} $\chi$ for all
the classes with $M=102$ (adapted from [45]).

{\bf Figure 15}: $\Delta^2(E)$ {\it vs} energy $E$ for different values of
$\chi$ for the $eo$ class with $M=102$ (adapted from [45]).

{\bf Figure 16}: Momenta $I_m$ {\it vs} energy $E$ for the $eo$ class with
$M=102$: (a) chaotic region, (b) quasi--integrable region
(adapted from [51]).

{\bf Figure 17}: Information measure $S$ {\it vs} energy $E$
for the $eo$ class with $M=102$:
(a) chaotic region, (b) quasi--integrable region (adapted from [51]).

{\bf Figure 18}: Transition probability $(ee)\to (oo)$ with $M=50$: (a)
regular region, (b) chaotic region (adapted from [53]).

{\bf Figure 19}: A schematic illustration of the average flow of $\gamma$-ray
decay from high-spin states induced by heavy-ion compound reactions.
The insert illustrates the spreading of the E2 transitions
(adapted from [55]).

{\bf Figure 20}: The rotational E2 strength distribution {\it vs}
the energy $E_{\gamma}$ for four initial levels with
different excitation energies above the yrast line (adapted from [57]).

{\bf Figure 21}: The probability distribution of the rotational E2
strengths $s_{ij}$ for $\gamma$-ray energies in the interval
$0.90 < E_{\gamma} < 1.05$ MeV. The dashed curve represents the Porter-Thomas
distribution. The excitation energy above the yrast line is indicated
for each bin (adapted from [57]).

{\bf Figure 22}: The rotational--strength distributions obtained
for transitions from 50 levels in each bin. The excitation
energy above the yrast line is indicated for each bin. The smoothed
distribution function is drawn with a solid curve. The transitions shown are
for $(-,1)$ in $^{168}Yb$ at $\hbar\omega =0.5$ MeV
(adapted from [58]).

{\bf Figure 23}: Mixing parameter $q$ {\it vs} the strength of the two-body
interaction $\Delta$, at two excitation energy intervals, $2.0-2.5$ MeV
(dashed line) and $3.0-3.5$ MeV (solid line) in superdeformed $^{152}Dy$,
and $0.5-1.0$ MeV (dashed line) and $1.5-2.0$ MeV
(solid line) in normal-deformed $^{168}Yb$.
In all cases $I^{\pi} =50^+$ (adapted from [59]).

{\bf Figure 24}: Average standard deviation of the E2-strength function
{\it vs} $\Delta$ for $50^+\to 48^+$ transitions emerging from the excitation
energy intervals $2.0-2.5$ MeV (dashed line) and $3.0-3.5$ MeV (solid line)
in superdeformed $^{152}Dy$, and $0.5-1.0$ MeV (dashed line) and
$1.5-2.0$ MeV (solid line) in normal-deformed $^{168}Yb$
(adapted from [59]).

{\bf Figure 25}: Mixing parameter $\alpha$ {\it vs} the continuum
coupling constant $\chi$ (adapted from [62]).

{\bf Figure 26}: Mean ratio of the width of the
broadest state to the total summed width of all complex eigenvalues
{\it vs} the continuum coupling constant $\chi$. Error bars
correspond to the spread over 50 matrices $160$x$160$
(adapted from [62]).

{\bf Figure 27}: Classical and quantal measures
of chaos {\it vs} $\nu$ ($\nu =0$: SU(3), $\nu =1$: U(5)).
Left: $\bar{\lambda}$ and $\sigma$ for angular momentum per boson $l=0.1$.
Right: $\omega$ and $\nu$ for the levels with $I=2^+$
(adapted from [67]).

\newpage

\begin{verse} {\large \bf References} \\
\vspace{0.6 cm}
[1] O. BOHIGAS and H.A. WEIDENM\"ULLER: {\it Ann. Rev. Nucl. Part.
Sci.}, {\bf 38}, 421 (1988) \\
\vspace{0.3 cm}
[2] P. MARMIER and E. SHELDON: {\it Physics of Nuclei and Particles},
vol. 2 (Academic Press, New York, 1970) \\
\vspace{0.3 cm}
[3] F. IACHELLO and A. ARIMA: {\it The
Interacting Boson Model} (Cambridge Univ. Press, Cambridge, 1987) \\
\vspace{0.3 cm}
[4] F. IACHELLO and P. VAN ISACKER: {\it The Interacting Boson--Fermion Model}
(Cambridge Univ. Press, Cambridge, 1991) \\
\vspace{0.3 cm}
[5] I. TALMI: {\it Simple Models of Complex Nuclei}
(Harwood Academic Press, Paris, 1993) \\
\vspace{0.3 cm}
[6] A. BOHR and B.R. MOTTELSON: {\it Nuclear Structure}, vol. 1
(Benjamin, London, 1969) \\
\vspace{0.3 cm}
[7] A. BOHR and B.R. MOTTELSON: {\it Nuclear Structure}, vol. 2
(Benjamin, London, 1975) \\
\vspace{0.3 cm}
[8] M.H. MACFARLANE: in {\it Proc.Int. School Enrico Fermi}, Course XL
(North--Holand, New York, 1969)\\
\vspace{0.3 cm}
[9] N. BOHR: {\it Nature}, {\bf 137}, 344 (1936) \\
\vspace{0.3 cm}
[10] C.E. PORTER: {\it Statistical Theories of Spectra: Fluctuations},
(Academic Press, 1965) \\
\vspace{0.3 cm}
[11] F. DYSON: {\it J. Math. Phys.} {\bf 3}, 140 (1962) \\
\vspace{0.3 cm}
[12] F. DYSON and M. MEHTA: {\it J. Math. Phys.} {\bf 4}, 701 (1963) \\
\vspace{0.3 cm}
[13] E. WIGNER: {\it SIAM Rev.} {\bf 9}, 1 (1967) \\
\vspace{0.3 cm}
[14] M. MEHTA: {\it Random Matrices}, II Ed. (Academic Press, 1991) \\
\vspace{0.3 cm}
[15] M. MEHTA: in {\it Statistical Properties of Nuclei},
Ed. by J. GARG (1972) \\
\vspace{0.3 cm}
[16] O. BOHIGAS: in {\it Chaos and Quantum Physics}, Les Houches 1989,
Ed. M.J. GIANNONI, A. VOROS, and J. ZINN--JUSTIN (North--Holland,
Amsterdam, 1991) \\
\vspace{0.3 cm}
[17] D. BISWASH, S. SINHA and S.V. LAWANDE: Phys. Rev. A {\bf 46}, 2649
(1992) \\
\vspace{0.3 cm}
[18] R.U. HAQ, A. PANDEY and O. BOHIGAS: {\it Phys. Rev. Lett.} {\bf 48},
1086 (1982) \\
\vspace{0.3 cm}
[19] O. BOHIGAS, R.U. HAQ and A. PANDEY: in {\it Nuclear Data for
Science and Technology}, Ed. by K.H. BOCKHOFF (1983) \\
\vspace{0.3 cm}
[20] J.D. GARRETT, J.R. GERMAN, L. COURTNEY and J.M. ESPINO:
in {\it Future Directions in Nuclear Physics}, pag. 345,
Ed. by J. DUDEK, B. HAAS (American Institute of Physics, New York,
1992); J.D. GARRETT, J.R. GERMAN: {\it preprint ORNL/94},
submitted to {\it Phys. Rev. Lett.}\\
\vspace{0.3 cm}
[21] J.F. SHRINER, G.E. MITCHELL and T. VON EGIDY:
{\it Z. Phys.} A {\bf 338}, 309 (1990) \\
\vspace{0.3 cm}
[22] V.I. ARNOLD: {\it Mathematical Methods of Classical Mechanics}
(Springer--Verlag, Berlin, 1978); V.I ARNOLD and A. AVEZ: {\it Ergodic
Problem of Classical Mechanics} (Benjamin, London, 1968) \\
\vspace{0.3 cm}
[23] M. TABOR: {\it Chaos and Integrability in Non Linear Dynamics},
(Wiley, New York, 1989) \\
\vspace{0.3 cm}
[24] O. DE ALMEIDA: {\it Hamiltonian Systems: Chaos and
Quantization} (Univ. Press, Cambridge, 1990) \\
\vspace{0.3 cm}
[25] M.C. GUTZWILLER: {\it Chaos in Classical and Quantum Mechanics}
(Springer-Verlag, Berlin, 1991)\\
\vspace{0.3 cm}
[26] YA.G. SINAI: {\it Introduction to Ergodic Theory} (Univ. Press,
Princeton, 1977) \\
\vspace{0.3 cm}
[27] C. COHEN--TANNOUDJI, B. DIN and F. LALOE: {\it Quantum
Mechanics} (1977) \\
\vspace{0.3 cm}
[28] O. BOHIGAS, M. GIANNONI and C. SCHMIT: {\it Phys. Rev. Lett.} {\bf 52}, 1
(1984) \\
\vspace{0.3 cm}
[29] O. BOHIGAS, M. GIANNONI: in {\it Mathematical and Computational
Methods in Nuclear Physics}, Ed. by J.S. DEHESA, J.M.G. GOMEZ and A.
POLLS, {\it Lecture Note in Physics}, vol. 209 (Springer, Berlin 1984) \\
\vspace{0.3 cm}
[30] M.V. BERRY and M. TABOR: {\it Proc. Roy. Soc. Lond.} A {\bf 356},
375 (1977) \\
\vspace{0.3 cm}
[31] A. PANDEY, O. BOHIGAS and M.J. GIANNONI: {\it J. Phys.} A {\bf 22},
4083 (1989); A. PANDEY and R. RAMASWAMY: {\it Phys. Rev.} A {\bf 43}, 4237
(1991)\\
\vspace{0.3 cm}
[32] M.C. GUTZWILLER: {\it J. Math. Phys.} {\bf 11}, 1791 (1970) \\
\vspace{0.3 cm}
[33] M.C. GUTZWILLER: {\it J. Math. Phys.} {\bf 12}, 343 (1971)  \\
\vspace{0.3 cm}
[34] M. BERRY: {\it Proc. Roy. Soc. Lond.} A {\bf 400}, 229 (1985) \\
\vspace{0.3 cm}
[35] H. LIPKIN, N. MESHKOV and A. GLICK: {\it Nucl. Phys.} {\bf 62}, 188
(1965) \\
\vspace{0.3 cm}
[36] R. WILLIAMS and S. KOONIN: {\it Nucl. Phys.} A {\bf 391}, 72 (1982)\\
\vspace{0.3 cm}
[37] D. MEREDITH, S. KOONIN and M. ZIRNBAUER: {\it Phys. Rev.} A
{\bf 37}, 3499 (1988) \\
\vspace{0.3 cm}
[38] R. GILMORE and H. FENG: {\it Prog. Part. and Nucl. Phys.}
{\bf 9}, 479 (1983) \\
\vspace{0.3 cm}
[39] SUBROUTINE D02BAF: {\it The NAG Fortran Library}, Mark 14, Oxford:
NAG Ltd. and USA: NAG Inc. (1990) \\
\vspace{0.3 cm}
[40] SUBROUTINE E04JAF: {\it The NAG Fortran Library}, Mark 14, Oxford:
NAG Ltd. and USA: NAG Inc. (1990) \\
\vspace{0.3 cm}
[41] M. BARANGER and K. DEVIS: {\it Ann. Phys.} {\bf 177}, 330 (1987);
M. DE AGUIAR, C. MALTA, M. BARANGER and K. DEVIS: {\it Ann. Phys.}
{\bf 180}, 167 (1987);
M. BARANGER, K. DEVIS and J. MAHONEY: {\it Ann. Phys.} {\bf 183},  95 (1988) \\
\vspace{0.3 cm}
[42] H.D. MEYER: {\it J. Chem. Phys.} {\bf 84}, 3147 (1986) \\
\vspace{0.3 cm}
[43] V.R. MANFREDI and L. SALASNICH: {\it Z. Phys.} A {\bf 343}, 1 (1992)  \\
\vspace{0.3 cm}
[44] S. $\AA$BERG: {\it Prog. Part. Nucl. Phys.} {\bf 28}, 11 (1992) \\
\vspace{0.3 cm}
[45] V.R. MANFREDI, L. SALASNICH and L. DEMATT\`E: {\it Phys. Rev.} E {\bf
47} 4556 (1993)\\
\vspace{0.3 cm}
[46] D.W. NOID, M. KOSZYKOWSKI, M. TABOR and R.A. MARCUS: {\it J. Chem.
Phys.}, {\bf 72}, 6169 (1980) \\
\vspace{0.3 cm}
[47] R. RAMASWAMY and R.A. MARCUS: {\it J. Chem. Phys.} {\bf 74}, 1385
(1991) \\
\vspace{0.3 cm}
[48] M.V. BERRY: in {\it Chaotic Behaviour in Quantum Mechanics}, pag.
121, Ed. by G. CASATI (Plenum, New York, 1984) \\
\vspace{0.3 cm}
[49] I.C. PERCIVAL: {\it Adv. Chem. Phys.}, {\bf 36}, 1 (1977)  \\
\vspace{0.3 cm}
[50] L. DEMATT\`E, V.R. MANFREDI and L. SALASNICH:
in {\it SIF Conference Proc. 'From Classical to Quantum Chaos'},
vol. 41, pag. 111, Ed. by G.F. DELL'ANTONIO, S. FANTONI,
V.R. MANFREDI (Editrice Compositori, Bologna, 1993) \\
\vspace{0.3 cm}
[51] M.T. LOPEZ--ARIAS and V.R. MANFREDI: {\it Z. Phys.} A {\bf 334} 255
(1989); M.T. LOPEZ--ARIAS, V.R. MANFREDI: in {\it 3rd Int. Spring Seminar
on Nuclear Physics} Ed. by A. COVELLO (1990) \\
\vspace{0.3 cm}
[52] M.V. BERRY: {\it J. Phys. A} {\bf 10}, 2083 (1977); M.V. BERRY,
J.H. HANNAY and A.M. OZORIO DE ALMEIDA: {\it Physica D} {\bf 8}, 229
(1983)\\
\vspace{0.3 cm}
[53] M.T. LOPEZ--ARIAS and V.R. MANFREDI: {\it Nuovo Cimento} A
{\bf 104}, 283 (1991) \\
\vspace{0.3 cm}
[54] D. BAZZACCO: in {\it Int. Conf. on Nuclear Structure at High
Angular Momenta}, vol. 2, pag. 376, AECL 10613, Ottawa (1992)\\
\vspace{0.3 cm}
[55] J.D. GARRETT, G.B. HAGEMANN and B. HERSKIND: {\it Ann. Rev. Nucl.
Part. Sci.} {\bf 36}, 419 (1986)\\
\vspace{0.3 cm}
[56] T. GUHR and H.A. WEIDENM\"ULLER: {\it Ann. Phys.} {\bf 193}, 472 (1989) \\
\vspace{0.3 cm}
[57] M. MATSUO, T. DOSSING, B. HERSKIND and S. FRAUENDORF:
{\it Nucl. Phys.} A {\bf 557}, 211c (1993)\\
\vspace{0.3 cm}
[58] M. MATSUO, T. DOSSING, B. HERSKIND and S. FRAUENDORF:
Nucl. Phys. A {\bf 564}, 345 (1993) \\
\vspace{0.3 cm}
[59] S. $\AA$BERG: {\it Phys. Rev. Lett.} {\bf 64}, 3119 (1990) \\
\vspace{0.3 cm}
[60] T.A. BRODY, J. FLORES, J.B. FRENCH, P.A. MELLO, A. PANDEY and S.S.
WONG: {\it Rev. Mod. Phys.} {\bf 53}, 385 (1981) \\
\vspace{0.3 cm}
[61] V.V. SOKOLOV and V.G. ZELEVINSKY: {\it Nucl. Phys.} A {\bf 504},
562 (1989) \\
\vspace{0.3 cm}
[62] S. MIZUTORI and V.G. ZELEVINSKY: {\it Z. Phys.} A {\bf 346}, 1 (1993) \\
\vspace{0.3 cm}
[63] J. GINIBRE: {\it J. Math. Phys.} {\bf 6}, 3 (1965) \\
\vspace{0.3 cm}
[64] F. HAAKE: {\it Quantum Signature of Chaos} (Springer, Berlin, 1991) \\
\vspace{0.3 cm}
[65] F. HAAKE, F. IZRAILEV, N. LEHMANN, D. SAHER and H.J. SOMMERS:
{\it Z. Phys. B} {\bf 88}, 359 (1992)\\
\vspace{0.3 cm}
[66] R. GROBE, F. HAAKE and H.J. SOMMERS: {\it Phys. Rev. Lett.} {\bf 61},
1899 (1988) \\
\vspace{0.3 cm}
[67] N.WHELAN and Y.ALHASSID: {\it Nucl. Phys.} A {\bf 556}, 42 (1993) \\

\end{verse}

\end{document}